# Energy Dissipation and Transport in Nanoscale Devices

Eric Pop

*Dept. of Electrical and Computer Engineering, Micro and Nanotechnology Lab and Beckman Institute*
*University of Illinois Urbana-Champaign, Urbana IL 61801, U.S.A.*
Contact: epop@illinois.edu

**ABSTRACT**

Understanding energy dissipation and transport in nanoscale structures is of great importance for the design of energy-efficient circuits and energy-conversion systems. This is also a rich domain for fundamental discoveries at the intersection of electron, lattice (phonon) and optical (photon) interactions. This review presents recent progress in understanding and manipulation of energy dissipation and transport in nanoscale solid-state structures. First, the landscape of power usage from nanoscale transistors ($\sim 10^{-8}$ W) to massive data centers ($\sim 10^{9}$ W) is surveyed. Then, focus is given to energy dissipation in nanoscale circuits, silicon transistors, carbon nanostructures, and semiconductor nanowires. Concepts of steady-state and transient thermal transport are also reviewed in the context of nanoscale devices with sub-nanosecond switching times. Finally, recent directions regarding energy transport are reviewed, including electrical and thermal conductivity of nanostructures, thermal rectification, and the role of ubiquitous material interfaces.

## 1. Introduction

Some of the greatest challenges of modern society are related to energy consumption, dissipation and waste. Among these, present and future technologies based on nanoscale materials and devices hold great potential for improved energy consumption, conversion, or harvesting. A prominent example is that of integrated electronics, where power dissipation issues have recently become one of its greatest challenges. Power dissipation limits the performance of electronics from mobile devices ($\sim 10^{-3}$ W) to massive data centers ($\sim 10^{9}$ W), all primarily based on silicon micro/nanotechnology. Put together, the energy use of the United States information technology (IT) infrastructure is currently in excess of 20 GW, or 5-10% of our national electricity budget, with an approximate breakdown as shown in Figs. 1c and 1d [1]. Importantly, the figures for data center energy consumption have doubled in the past five years, with waste heat requiring drastic cooling solutions (Fig. 1c). Such challenges are also evident at the individual microprocessor (CPU) level, where the race to increase operating frequency beyond a few GHz recently stopped when typical dissipated power reached 100 W/cm$^2$ (Fig. 1b), an order of magnitude higher than a typical hot plate [2]. Such electronic power and thermal challenges have negative impacts from massive database servers to new applications like wearable devices, medical instrumentation, or portable electronics. In the latter situations, a basic trade-off is one between the available functionality and the need to carry heavy batteries to power it.

Despite tremendous progress over the past three decades, modern silicon transistors are still over three orders of magnitude (>1000×) more energy inefficient than fundamental physical limits, as shown in Fig. 1a. These limits have been estimated as approximately $3k_BT \approx 10^{-20}$ J at room temperature for a binary switch with a single electron and energy level separation $k_BT$, where $k_B$ is the Boltzmann constant and $T$ is the absolute temperature [3]. In the average modern microprocessor the dissipated power is due, in approximately equal parts, to both leakage (or "sleep") power and active (dynamic) switching power [4], as detailed in Section 2. Power dissipation is compounded at the system level, where each CPU Watt demands approximately 1.5x more for the supply, PC board, and case cooling [1]. Such power (mis)use is even more evident in systems built on otherwise power-efficient processors, e.g. in the case of the Intel Atom N270 (2.5 W power use) which is typically paired up with the Intel 945GSE chipset (11.8 W power use) [5]. At the other extreme, data centers require 50-100% additional energy for cooling (Fig. 1c), which is now the most important factor limiting their performance, not the hardware itself.

If present growth trends are maintained, data center and overall electronics power use could reach one third of total US consumption by 2025 [1]. Worldwide, the growth trends could be even steeper, given that technolo-gically developed regions such as the US, Western Europe and Japan currently account for 58 percent of the world's computers, but only 15 percent of the world population. [6]. Such energy challenges for the electronics infrastructure stem not only from the power supply side which call for new energy sources, efficient batteries, or thermoelectrics, but also from the demand side, i.e. the need for more energy-efficient computing devices. To put it in financial terms, data centers consumed more than 7 GW or $4.5 billion in 2006 (Fig. 1c) [1]. The estimated 108 million PCs in offices across the US had an annual energy cost of $4.2 billion in 2008 [6]. More troubling, the amount of equivalent $CO_2$ emissions generated is approximately equivalent to that of 5 million cars, comparable to the entire state of Maryland.

In fact, such estimates are likely to be conservative, missing the power consumed by home PCs, routers, networks, and the Internet backbone where detailed studies are not available (however, a thought-provoking study of energy consumption by e-mail SPAM has recently been made [8]). Nevertheless, the overall negative impact of these trends (to predicted 2025 levels) on energy supplies, budgets, and the environment in terms of equivalent $CO_2$ emissions is staggering. Thus, breakthroughs in our understanding and improvement of energy efficiency in nanoelectronics will also have a global effect, impacting the entire structure of modern society.

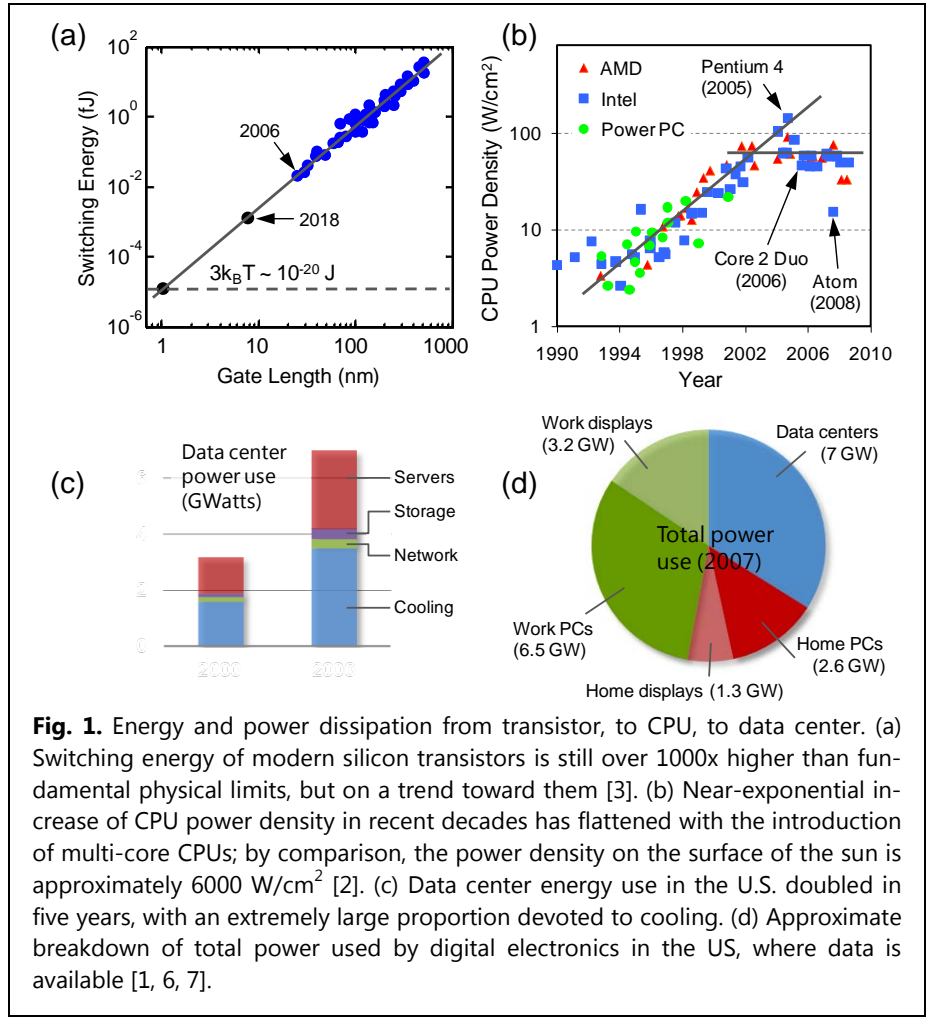

**Fig. 1.** Energy and power dissipation from transistor, to CPU, to data center. (a) Switching energy of modern silicon transistors is still over 1000x higher than fundamental physical limits, but on a trend toward them [3]. (b) Near-exponential increase of CPU power density in recent decades has flattened with the introduction of multi-core CPUs; by comparison, the power density on the surface of the sun is approximately 6000 W/cm$^2$ [2]. (c) Data center energy use in the U.S. doubled in five years, with an extremely large proportion devoted to cooling. (d) Approximate breakdown of total power used by digital electronics in the US, where data is available [1, 6, 7].

On a broader scale, just over half the man-made energy in the world is wasted as heat ($10^{13}$ W), from power plants and factories, to car engines and the power bricks on our laptops. Efficiently reclaiming even a small percentage of such wasted heat would itself nearly satisfy the electricity needs of our planet [7]. The fundamental issues at hand are, in fact, a two-sided problem: On one side, there is a significant need for low-energy computing devices, which is perhaps the biggest challenge in micro/nanoelectronics today. On the other side lies the challenge of waste heat dissipation, guiding, or conversion into useful electricity.

This review discusses several aspects of the above, from the nanoscale circuits, device and materials perspective, as follows: Section 2 examines energy dissipation and optimization in nanoscale circuits, with focus on leakage vs. active power in nanoscale technologies. Section 3 presents energy dissipation in nanoscale devices (e.g. silicon transistors) from the diffusive to the ballistic regime. Section 4 reviews concepts of steady-state and transient thermal transport in the context of nanoscale devices with sub-nanosecond switching times. Finally, recent directions regarding energy transport are reviewed, including electrical and thermal conductivity of carbon nanostructures and nanowires (Section 5), thermal rectification (Section 6), and the role of ubiquitous material interfaces (Section 7).

## 2. Energy Dissipation in Nanoscale Circuits

Energy and power dissipation in nanoscale digital circuits is often described in the context of inverter activity, as shown in Fig. 2. Three components have been traditionally identified for digital power consumption: dynamic power used during switching for charging and discharging the inverter load, sub-threshold leakage power, and short-circuit power [9]:

$$P = C_L V_{DD}^2 \alpha f + I_{leak} V_{DD} + P_{SC} \quad (1)$$

where $\alpha$ is the activity factor, $C_L$ is the load capacitance, $f$ is the clock frequency and $I_{leak}$ is the sub-threshold leakage current [10]. The short-circuit power is typically the smallest for well-designed circuits with equal rise and fall times [9, 11]. The leakage component, however, is a strong function of temperature $T$, and therefore implicitly dependent on the total power dissipated. Moreover, leakage current also scales exponentially with the voltage overdrive ($V_{GS} - V_T$) and is very sensitive to changes in threshold voltage:

$$I_{leak} = \frac{W}{L_{eff}} \mu_{eff} C_{OX} (m-1) \left(\frac{k_B T}{q}\right)^2 \exp\left(\frac{V_{GS} - V_T}{m k_B T/q}\right) \left[1 - \exp\left(-\frac{V_{DS}}{k_B T/q}\right)\right] \quad (2)$$

where $\mu_{eff}$ is the effective mobility, $C_{OX}$ the gate oxide capacitance, $W$ and $L_{eff}$ the effective channel width and length, $T$ the absolute temperature, $V_{DS}$ the drain voltage, $V_{GS}$ the gate voltage, $V_T$ the threshold voltage, and $m$ the subthreshold body factor [11]. With (down)scaling of technology dimensions and voltage, the role of sub-threshold leakage has become increasingly important, as shown by the trends in Fig. 2b and 2c. This is due to a reduction in supply voltage $V_{DD}$ without the ability to properly scale the threshold voltage, in addition to short-channel phenomena like Drain Induced Barrier Lowering (DIBL) [9].

An interesting trade-off becomes apparent if we compare the energy (rather than the power) dissipated by a simple digital circuit like an inverter chain, as shown in Fig. 2b, here for a 130 nm technology with threshold voltage $V_T = 0.4$ V. Energy per operation is the more important metric for applications that are limited by battery life, such as mobile or medical devices. The dynamic energy, like the dynamic power, scales quadratically with the supply voltage $V_{DD}$. However, the leakage energy ($E_{leak} = I_{leak} V_{DD} t_d$) rises sharply as the supply voltage is reduced because the circuit delay ($t_d$) increases in sub-threshold ($V_{DD} \leq V_T$) operation. This raises an interesting point: for applications that are not concerned with maximum performance, the energy-optimal operation occurs in the sub-$V_T$ regime for a given CMOS technology, as shown in Fig. 2b [10, 11].

In applications that do seek optimal performance (such as desktop or server CPUs), leakage power dissipation has also become significant, as illustrated in Fig. 2c. Particularly with respect to nanoscale devices with picosecond switching delays, Eq. (2) above merits a closer look, upon which several of its shortcomings become evident. First, the expression was derived for 3-dimensional diffusive transport across an energy barrier whereas most nanoscale devices are 1- or 2-dimensional (the leakage current of a carbon nanotube transistor was recently derived in Ref. [12]). Second, leakage is strongly dependent on temperature, yet this itself is highly unsteady during transient digital switching. Typical electrical transients and inverter delays are of the order 1-10 ps, whereas device thermal time constants are ~10 ns [13, 14]. Which temperature is used in computing the leakage current is an important choice, and using an average "junction temperature" as is often done [15] can lead to significant over- or under-estimates of the total leakage. Third, even if the thermal transients are properly accounted for in evaluating leakage during digital operation, the notion of temperature itself may need to be reconsidered when switching approaches the time scales of electron-phonon and phonon-phonon collisions (0.1-10 ps), leading to non-equilibrium conditions. Electrons scatter strongly with optical phonons (OP) in all materials and devices under consideration, and OP lifetimes are themselves of the order 1-10 ps [16-20]. Optical phonons range from approximately 35 meV in Ge and GaAs, to 60 meV in silicon, and nearly 200 meV in carbon nanotubes and graphene [21], and hence the absorption of a single OP can lead to enough additional energy to immediately surpass the potential barrier in a MOSFET-like device (Fig. 3c). Recent results have also shown significant increase in band-to-band tunneling currents in carbon nanotubes due to optical phonon absorption [22]. As will be seen below, it is apparent that our present-day understanding and modeling of power dissipation in nanoscale circuits falls short of the recent advances made in nanoscale devices. This remains an area where significant progress remains to be made.

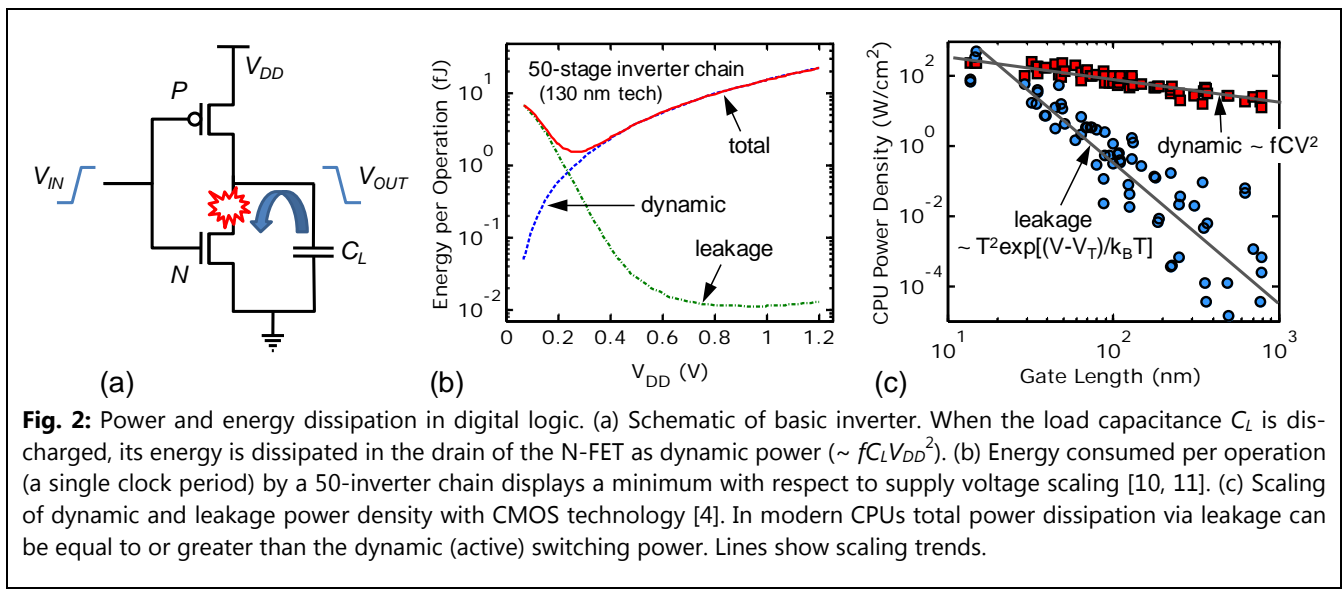

**Fig. 2:** Power and energy dissipation in digital logic. (a) Schematic of basic inverter. When the load capacitance $C_L$ is discharged, its energy is dissipated in the drain of the N-FET as dynamic power (~ $fC_L V_{DD}^2$). (b) Energy consumed per operation (a single clock period) by a 50-inverter chain displays a minimum with respect to supply voltage scaling [10, 11]. (c) Scaling of dynamic and leakage power density with CMOS technology [4]. In modern CPUs total power dissipation via leakage can be equal to or greater than the dynamic (active) switching power. Lines show scaling trends.

## **3. Energy Dissipation in Nanoscale Devices**

### *3.1 Diffusive, Ballistic, and Contacts*

The most elementary approach for estimating the power dissipation of an electronic device is to express it as the product of the current passing through and the voltage drop across it:

$$P_D = I \times V = I^2 R . \tag{3}$$

This is the classical expression for a device under diffusive transport (subscript $D$), and the voltage drop excludes the device contacts (Fig. 3a). Hence, this equation must be applied with care in describing the power dissipated in a structure with relatively large contact resistance ($R_C$), e.g. a carbon nanotube or molecular device. The intrinsic device resistance also excludes the quantum contact resistance $R_0 = h/(Mq^2) \approx 25.8/M$ k$\Omega$, where $q$ is the elementary charge, $h$ is Planck's constant and $M$ is the number of modes for electrical transport (e.g., $M = 4$ for single-wall carbon nanotubes, accounting for band degeneracy and spin) [23-26]. In large classical devices this can be ignored, as the presence of many transport modes ensures that $R_0$ is extremely small.

The expression above will overestimate the total power dissipated in a quasi-ballistic device, i.e. one with dimensions comparable to or shorter than the inelastic scattering length [27, 28], as schematically shown in Fig. 3b. In this case, carriers gain energy comparable to the applied voltage ($E \sim qV$) but do not undergo enough inelastic scattering events to equilibrate and dissipate it to the lattice by the time they exit. In order to estimate dissipation in quasi-ballistic devices (subscript $QB$), we introduce the inelastic mean free path (MFP) $\lambda_{OP}$, typical for carrier scattering with high-energy optical phonons ($\hbar\omega_{OP}$):

$$P_{QB} = \frac{\hbar\omega_{OP}}{q} \left( \frac{L}{L+\lambda_{OP}} \right) \frac{V}{R_0} \tag{4}$$

where $R_0$ is the quantum of electrical resistance defined above, $\lambda_{OP} = L(\hbar\omega_{OP}/qV) + \lambda_0$, and $\lambda_0$ is the spontaneous OP emission MFP for carriers with energy $E > \hbar\omega_{OP}$ [29, 30] (for example, $\lambda_0 \approx 15d$ for a carbon nanotube with diameter $d$ [31]). Here it is assumed that the current flowing through the device is limited by the inelastic scattering length, $I = (V/R_0)\lambda_{OP}/(L+\lambda_{OP})$ and that a carrier undergoes approximately $L/\lambda_{OP}$ inelastic collisions within the device, each time losing an energy $\hbar\omega_{OP}$ to the lattice. In addition, the carrier is assumed to relax to the bottom of its band immediately after the inelastic phonon emission event (Fig. 3b). Near the limit of zero temperature, an even sharper onset of power dissipation is expected, as soon as the applied voltage is large enough for carriers to emit a single phonon, $V \geq \hbar\omega_{OP}/q$, as shown for molecular junctions [32-35]. The expression above recovers the correct limit of power dissipation within a device in the ballistic limit ($P \to 0$ as $L \ll \lambda_{OP}$), although a more careful examination reveals that the scattering length itself depends on the phonon occupation and tem-perature at sufficiently high power levels. In this case, the electrical-thermal dissipation must be self-consistently solved, as in Refs. [30, 36]. Moreover, in quasi-ballistic transport hot electrons (holes) escape through the contact at higher (lower) voltage and a significant portion of the total power is dissipated there instead [37]. This is interpreted as the difference between the input power and the power dissipated within the quasi-ballistic device ($P_D$-$P_{QB}$ above).

A third example concerns heat dissipation during thermionic or tunneling transport across mesoscopic energy barriers, which are often encountered in transistors, diodes, heterojunctions and device contacts (Fig. 3c). Such dissipative transport does not lend itself to a simple, analytic interpretation, but it has been examined with the non-equilibrium Green's functions (NEGF) approach [27, 38], by Monte Carlo simulations [39], and the scattering matrix method [40]. A key feature of dissipation near barriers is the role of the energy barrier as a filter of the carrier distribution function, as schematically shown in Fig. 3c. Carriers with higher energy have a greater transmission probability (either thermionic or tunneling) across the barrier, leading to a "cooler" carrier distribution to the left (uphill) of the barrier, and a "hotter" distribution to the right (downhill). This results in an effective cooling (negative heat dissipation) to the left, and heating of the lattice (positive heat dissipation) to the right of the barrier. This effect depends not only on the barrier shape and height, but also on the direction of the current flow. In addition, such asymmetric heating can also be observed at metal-semiconductor contacts [41, 42], at contact energy barriers between dissimilar semiconductors [43-45], and even at more subtle transitions between regions with different density of states (DOS). The latter is a thermoelectric effect recently observed in transport between monolayer and bilayer graphene [46] and is a result of hot carriers diffusing from regions of lower to higher DOS in order to maximize the entropy (also equivalent to the classical effect of a gas that cools as it expands).

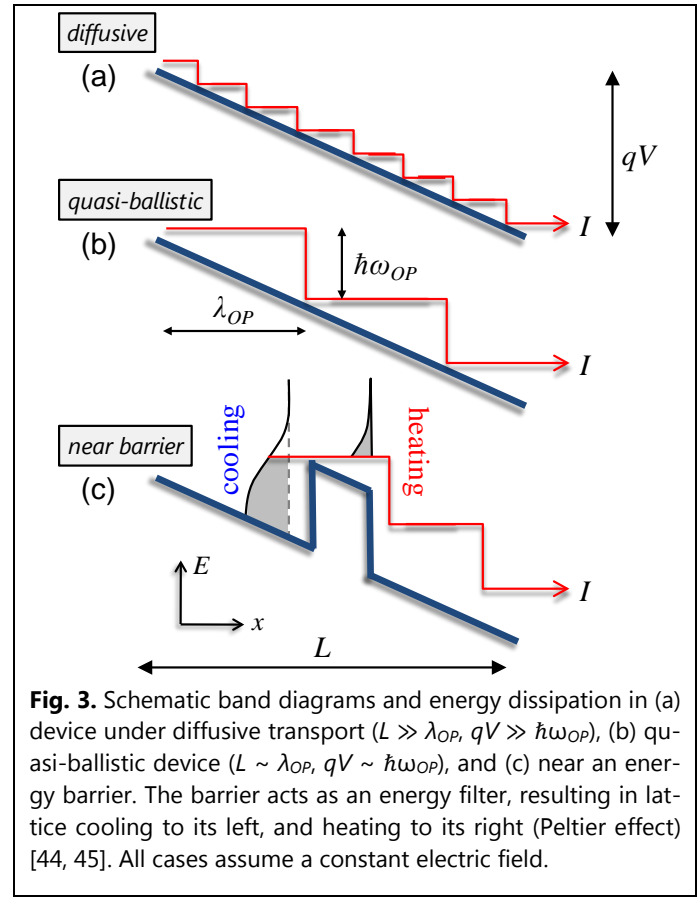

**Fig. 3.** Schematic band diagrams and energy dissipation in (a) device under diffusive transport ($L \gg \lambda_{OP}$, $qV \gg \hbar\omega_{OP}$), (b) quasi-ballistic device ($L \sim \lambda_{OP}$, $qV \sim \hbar\omega_{OP}$), and (c) near an energy barrier. The barrier acts as an energy filter, resulting in lattice cooling to its left, and heating to its right (Peltier effect) [44, 45]. All cases assume a constant electric field.

### *3.2 Spatial Distribution of Energy Dissipation*

Returning in more detail to the case of diffusive transport (dimension $L \gg$ inelastic MFP), more sophisticated methods are needed to compute the *spatial* distribution of power dissipation, rather than the simple lumped result of Eq. 3. This is most often given by the drift-diffusion approach [47-50]:

$$P_V = \mathbf{J} \cdot \mathbf{E} + (R - G)(E_G + 3k_B T) \tag{5}$$

where **J** is the current density, **E** the electric field, ($R$-$G$) is the net non-radiative recombination rate (recombination minus generation), $E_G$ is the semiconductor band gap and $T$ is the lattice temperature. The factor $3k_BT$ arises if the average energy of an electron (hole) above (below) the conduction (valence) band is ~$3/2k_BT$. The result above is typically implemented as a finite-element simulation on a device grid. Note the notation of $P_V$ (power density per unit volume, i.e. W/m$^3$) vs. Eq. 3 above (total power in Watts). The total power dissipated can be recovered by integrating Eq. 5 over the device volume. The first term represents the Joule heating rate which is usually positive (power generation) as charge carriers move along the band structure under the influence of the electric field, giving up energy to the lattice. This equation may include higher order terms accounting for carrier drift along a temperature gradient or across a discontinuity in the band structure [43, 48].

Unfortunately, this field-dependent method does not account for the microscopic nature of heat generation near a strongly peaked electric field region, such as in the drain of the transistor. Although electrons gain most of their energy at the location of the peak field, they travel several mean free paths before releasing it to the lattice, in decrements of (at most) the optical phonon energy. In silicon, the optical phonon energy is $\hbar\omega_{OP} \approx 60$ meV and in carbon nanotubes or graphene it is approximately three times greater. Typical inelastic scattering mean free paths in both silicon and carbon nanostructures are of the order $\lambda_{OP} \approx 10\text{-}50$ nm [31]. The full electron energy relaxation length is thus even longer, i.e. several inelastic mean free paths. This is illustrated near a 20 nm wide energy barrier in silicon in Fig. 4, where the drift-diffusion approach cannot capture the delocalized nature of the power dissipation region.

An improvement is provided by the hydrodynamic approach, which introduces the electron temperature ($T_e$) and an average electron energy relaxation time ($\tau_{e-L}$) [48]:

$$P_V = \frac{3}{2}\frac{k_B n(T_e - T_L)}{\tau_{e-L}} + (R-G)\left[E_G + \frac{3k_B}{2}(T_e + T_L)\right] \quad (6)$$

where $n$ is the electron density and $L$ denotes the lattice. The equation here is written for electrons as majority carriers, but the holes can be treated similarly. Unlike the drift-diffusion model, this approach is better suited for capturing transport near highly peaked electric fields. However, this suffers from the simplification of a single averaged carrier temperature and relaxation time, as scattering rates are strongly energy dependent [51]. Neither method gives information regarding the frequencies and wave vectors of phonons emitted. Such details are important because the emitted phonons have different velocities and widely varying contributions to heat transport [52-55] and device heating [56, 57].

The mechanism by which lattice self-heating occurs is that of electron scattering with phonons, and therefore a model which deliberately incorporates all scattering events will also capture such energy dissipation details. Thus, the Monte Carlo (MC) method [59] originally developed for studying hot electron effects [60], is also well-suited for computing a detailed picture of energy dissipation. This was the approach adopted in Refs. [58, 61-64], where power dissipation was computed as a sum of all phonon emission minus all phonon absorption events:

$$P_V = \frac{n}{N_{sim}\Delta t}\sum(\hbar\omega_{ems} - \hbar\omega_{abs}) \quad (7)$$

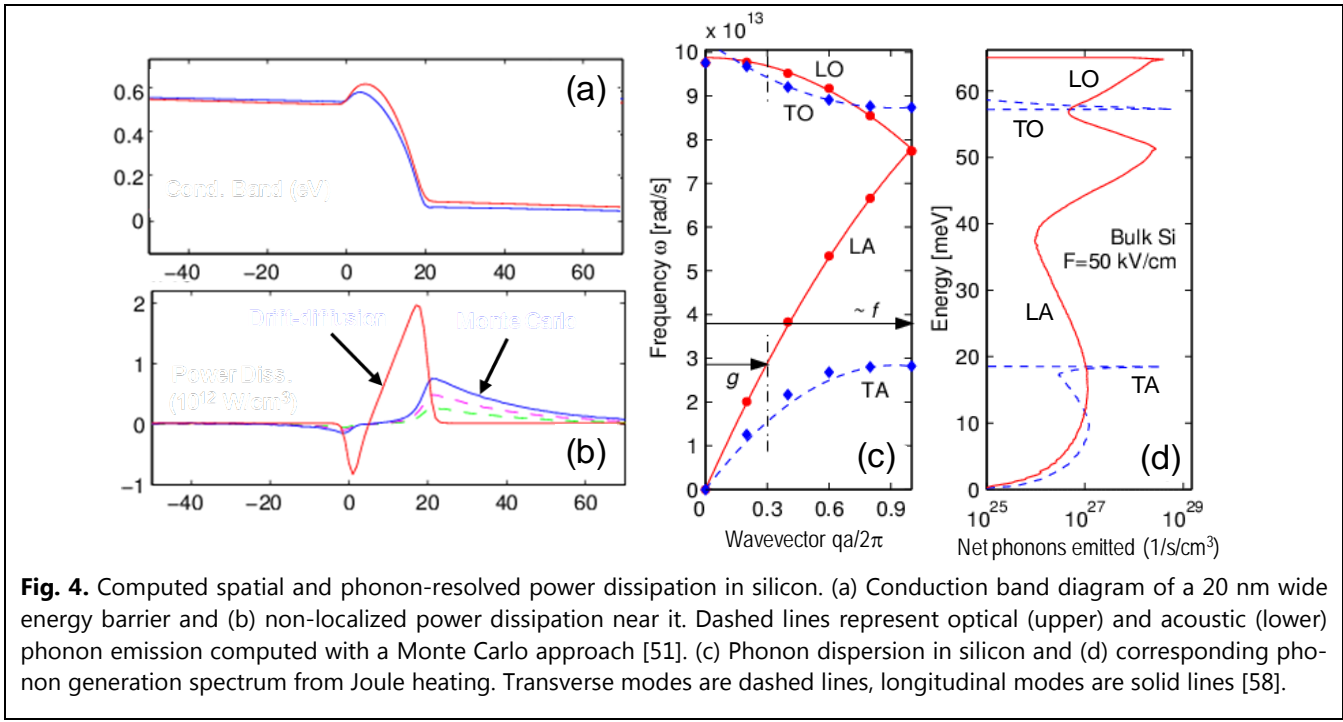

**Fig. 4.** Computed spatial and phonon-resolved power dissipation in silicon. (a) Conduction band diagram of a 20 nm wide energy barrier and (b) non-localized power dissipation near it. Dashed lines represent optical (upper) and acoustic (lower) phonon emission computed with a Monte Carlo approach [51]. (c) Phonon dispersion in silicon and (d) corresponding phonon generation spectrum from Joule heating. Transverse modes are dashed lines, longitudinal modes are solid lines [58].

where *n* is the real-space carrier density, $N_{sim}$ is the number of simulated particles (e.g. 10,000 simulated particles could be used to describe $10^{19}$ cm$^{-3}$ real-space concentration) and $\Delta t$ is the time. This approach has been used to investigate phonon emission as a function of frequency and mode in silicon, as well as to study heat generation near a strongly peaked electric field in a realistic device geometry. Fig. 4b shows the comparison between a classical drift-diffusion calculation and that of the MC result. The former tends to overestimate the peak heat generation rate and predicts a narrower heat generation region, which follows the shape of the electric field. The MC method suggests a broader heat generation domain extending inside the device drain, and limited by the electron-phonon scattering rate there.

The Monte Carlo approach also shows that heat generation in silicon is not evenly divided among phonon modes, but that acoustic phonon modes receive approximately 1/3 and optical phonons 2/3 of the energy dissipated, as shown in Fig. 4d. More specifically, the longitudinal optical (LO) *g*-type phonon is responsible for approximately 60% of the total energy dissipation [58]. Optical phonons have group velocities below 1000 m/s, and are thus much slower than the Brillouin zone center acoustic phonons typically responsible for heat transport in silicon (group velocity 5000-9000 m/s). This non-equilibrium phonon generation can lead to an energy transfer bottleneck [56]. In other words, a significant non-equilibrium phonon population may build up, particularly for the *g*-type LO mode. The generation rates for the other phonon modes are either smaller or their density of states (DOS) is larger (the DOS is proportional to the square of the phonon wave vector, which is largest at the edge of the Brillouin zone) and non-equilibrium effects are less significant.

The self-consistent study of Rowlette and Goodson [61] built on previous work by Pop [58] and Sinha [62], and coupled the Monte Carlo method for electron transport with a Boltzmann transport equation (BTE) approach for phonon transport. Rowlette found that while extremely high heat generation densities do exist in modern silicon devices (>$10^{12}$ W/cm$^3$), non-equilibrium hot phonon effects are strongly dependent on the phonon relaxation time. Available data points at an optical phonon lifetime $\tau \approx$ 0.4-2 ps for optical phonons near the Brillouin zone center [18], but the phonons involved in electron scattering are the *g*- and *f*-type shown in Fig. 4c, for which no experimental lifetimes are available. The discrepancy between the LO phonon temperature computed with the self-consistent vs. the classical approach can be as high as 250% in the device drain, near the location of highest electric field [61]. However, the electron mobility is typically determined by the temperature and scattering mechanisms near the source of the channel in such quasi-ballistic devices [65], and does not appear to be strongly affected by the additional LO population. Nevertheless, device reliability may ultimately be affected by such hot phonons, as suggested by a recent experimental study [66].

## 4. Thermal Transport in Nanoscale Devices

### *4.1 Steady-State Thermal Resistance*

While the previous sections focused on power dissipation in nanoscale devices and circuits, this section and the subsequent ones shift our attention to heat transport in and thermal spreading from nanoscale devices and materials. At the simplest level, heat dissipation from a lumped electronic device can be quantified by measuring its thermal resistance ($R_{TH}$) to the environment. This yields an average temperature rise of the device as:

$$\Delta T = PR_{TH} \qquad (8)$$

where *P* is the power (heat) dissipated in Watts. Note this is extremely similar to the electrical Ohm's law ($\Delta V = IR$), with temperature taking the role of voltage and power having the role of electric current flow. For instance, the thermal resistance to heat flow across a layer of thickness *t* and cross-sectional area *A* is written analogously to its electrical resistance, as $R_{TH} = t/(kA)$, where *k* is the material thermal conductivity. As with the electrical conductance (resistance) quantum, there also exists an equivalent ballistic quantum of thermal conductance, which is directly proportional to absolute temperature, $G_0 = \pi^2 k_B^2 T/(3h)$ or approximately 0.28 nW/K near 300 K [84, 85]. By comparison, the thermal conductance of an individual single-wall carbon nanotube at room temperature has been measured to be 2.4 nW/K [80].

In practice, nanoscale and semiconductor devices take more varied shapes (Fig. 5), have many modes available for thermal transport, and interfaces or 3-dimensional heat spreading make their analysis more difficult. The thermal resistance of many semiconductor devices has been measured through noise thermometry [67], gate electrode electrical resistance thermometry [73, 76], pulsed voltage measurements [13, 69], or an AC con-ductance method [14, 70, 72, 74]. Fig. 5a presents a summary of such experimental data produced in the literature over the past nearly two decades, covering a wide range of device dimensions and technologies. A clear trend emerges, showing that device thermal resistance increases as a power law with reduced device dimensions, and is reaching values well above $10^5$ K/W for device dimensions below 100 nm. Carbon nanotube devices present a particularly interesting case, showing relatively high thermal resistance. This is due partly because they are usually measured on a relatively thick $SiO_2$ layer (50-300 nm), partly because of their much smaller "footprint" area (CNT diameter $d$ ≪ any other typical device widths $W$), and partly due to the thermal resistance at the CNT-$SiO_2$ interface [30, 78, 86, 87]. In other words, it is clear that both geometry, thermal conductivity of materials, and the thermal resistance of material interfaces play a significant role in determining the thermal impedance of a nanoscale device.

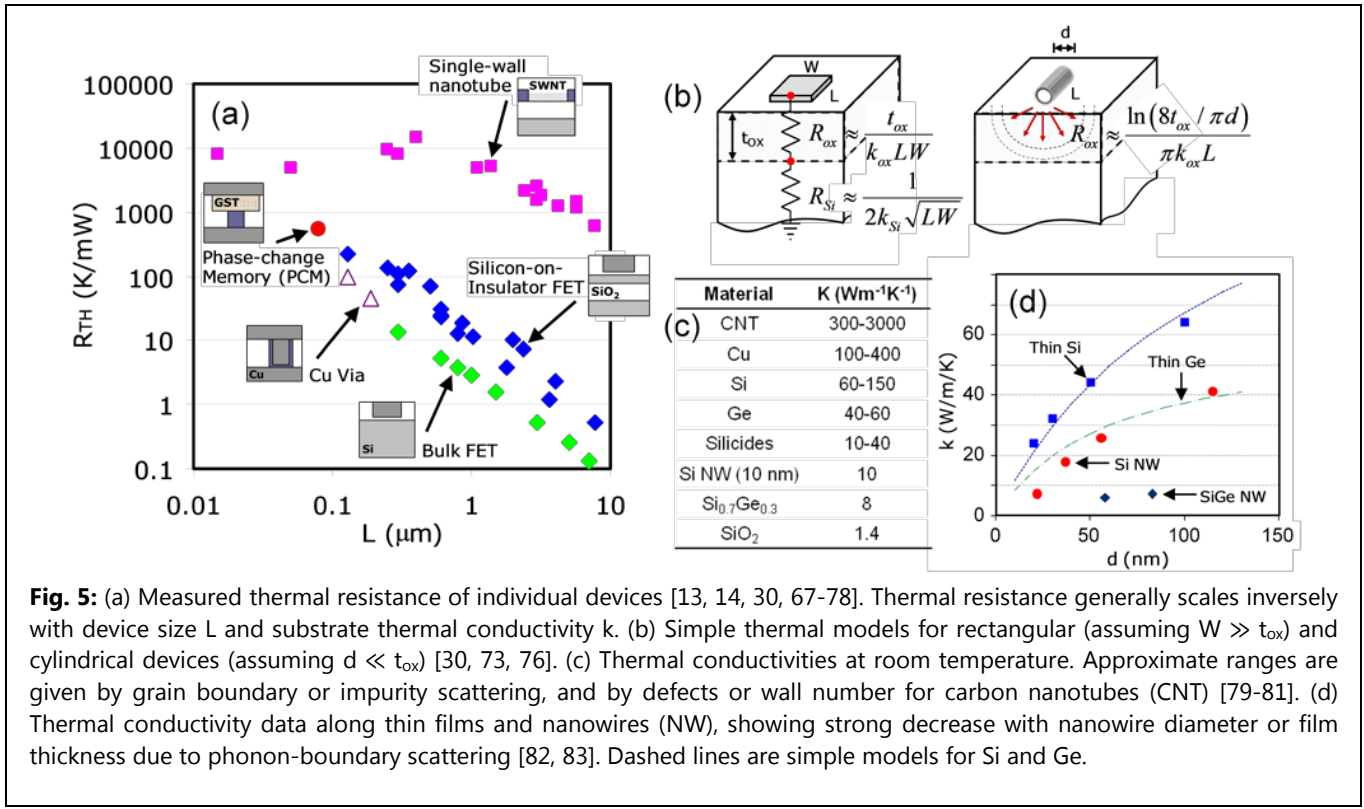

**Fig. 5:** (a) Measured thermal resistance of individual devices [13, 14, 30, 67-78]. Thermal resistance generally scales inversely with device size L and substrate thermal conductivity k. (b) Simple thermal models for rectangular (assuming W ≫ $t_{ox}$) and cylindrical devices (assuming d ≪ $t_{ox}$) [30, 73, 76]. (c) Thermal conductivities at room temperature. Approximate ranges are given by grain boundary or impurity scattering, and by defects or wall number for carbon nanotubes (CNT) [79-81]. (d) Thermal conductivity data along thin films and nanowires (NW), showing strong decrease with nanowire diameter or film thickness due to phonon-boundary scattering [82, 83]. Dashed lines are simple models for Si and Ge.

To understand the scaling displayed in Fig. 5a, consider the simplest case of a circular device with diameter $D$ on the surface of a silicon wafer (of thickness ≫ $D$). Its thermal resistance will scale as $R_{TH} = 1/(2k_{Si}D)$ as that of the spreading electrical resistance from a single circular contact [88]. This can be extended to a rectangular heat source (width $W$, length $L$) by replacing $D \approx (LW)^{1/2}$ and including additional 3-dimensional heat spreading shape factors [71, 73]. This case is illustrated in Fig. 5b, along with that of a cylindrical device (e.g. nanotube, nanowire). Several online tools are also available for quick, web-based spreading resistance calculations for various shapes and substrates [89]. Many other models of varying sophistication have been published, all of which reveal various inverse length and width dependencies [68, 73, 75, 76]. Naturally, the choice of such a model in practice depends on its complexity, and on the specific geometry of the device. Care must also be taken with the limiting cases $W \gg t_{ox}$ (1-dimensional heat flow through underlying oxide) as opposed to $d$ or $W \ll t_{ox}$ (3-dimensional spreading into oxide), the latter being the case for all nanotube, nanowire, and even graphene nanoribbon data [90] typically available. Finally, the thermal boundary resistance (TBR) at the interfaces between the devices and its environment can also be a limiting factor, as is the case with nanotubes [30, 91], phase-change memory devices [92, 93], and to some extent graphene transistors [94, 95]. The TBR in the latter case is approximately equivalent to the thermal resistance of 20-50 nm $SiO_2$ at room temperature, and of relatively little contribution to the commonly used graphene samples on 300 nm $SiO_2$. However, to estimate heat dissipation from graphene with thinner dielectrics, the TBR must be included in series with the additional material thermal resis-tances in Fig. 5b, at the interface between the device and its adjacent dielectric. More discussion on the TBR is provided in Section 7.

## *4.2 Transient Heat Conduction*

The thermal resistance models mentioned above are sufficient for evaluating the *steady-state* behavior of nanoscale devices, relevant during *I-V* characterization, analog operation, or to estimate the average temperature rise owed to device leakage. However, an understanding of *transient* heat conduction is necessary for short duration pulsed operation, such as during digital switching [9] or electrostatic discharge (ESD) events [96].

To first order, the temperature rise of a pulse-heated volume can be obtained from the energy of the heating pulse (*E*) and the heat capacity (*C*) of the volume (*V*) being heated:

$$\Delta T(t_p) \approx \frac{E}{CV} = \frac{Pt_p}{CV} \tag{9}$$

where $t_p$ is the pulse duration and *P* is its power. For instance, during digital operation the duration of an inverter switching event is approximately $t_p \approx$ 50-100 ps, which is significantly shorter than the thermal time constant of most modern devices, $\tau \approx$ 10-100 ns [97]. This is near the "adiabatic limit," where the device can be thermally decoupled from its environment. In other words, while the device is ON there is little spread of the heated volume outside the area where the actual heating takes place (here, the channel and drain of the transistor).

The approximation above only holds for heating pulses that are short enough not to cause any significant heating outside the device volume. For a pulse of duration $t_p$ the temperature spread extends approximately $(\pi \alpha t_p)^{1/2}$ outside the directly heated device volume [98], where $\alpha = k/C$ is the heat diffusion coefficient and *k* is the thermal conductivity (Fig. 5c). For heating pulses of $t_p =$ 50 ps this distance is of the order ~100 nm into silicon or germanium, and ~10 nm into adjacent $SiO_2$ layers (like the top passivation layer, or the buried oxide below silicon-on-insulator, i.e. SOI technology). For longer heating pulses $t_p \gg$ 50 ps or device dimensions *D* < 100 nm, we must take into account both the energy stored as temperature rise within the device and that in the surrounding dielectric (the heated volume of which expands as $\sim t_p^{1/2}$). In this case we can estimate [15, 99]:

$$\Delta T(t_p) \approx \frac{P}{CV}\left(2\sqrt{\tau t_p} - \tau\right) \tag{10}$$

where $t_p > \tau$ and $\tau = D^2/4\pi\alpha$ is the time constant associated with the minimum device dimension *D*. The expression above assumes the same heat capacity for both the device and its surrounding dielectric, which is a good approximation for Si and $SiO_2$ near room temperature ($C_{si} \approx C_{ox} \approx$ 1.7 Jcm$^{-3}$K$^{-1}$). Typical estimates of the temperature rise during digital switching have shown this dynamic value does not exceed a few degrees (e.g., 5 K) for sub-micron device technologies [14, 97]. However, the exact value for devices in the 10 nm range is unknown, and will be highly dependent on the ultimate choice of device geometry, materials and interfaces.

For longer time scales, comparable to or larger than the device thermal time constants, several models have been proposed [71, 75, 93, 99, 100]. These bridge the time range from the adiabatic limit to the steady-state operation of a device, and are typically based on a Green's functions solution of the heat diffusion equation. This approach is faster and offers more physical insight than solutions based on finite-element (FE) solvers. The disadvantage of such methods vs. the FE approach is their applicability to only a limited range of geometries, like heated sphere, infinite cylinder or rectangular parallelepiped, not taking into account the full geometry and diverse materials making up a modern semiconductor device.

Finally, for time scales or power pulses longer than the thermal diffusion time through the silicon wafer backside (~0.3 ms assuming a 500 μm thick wafer) the problem returns to a simpler one of steady-state temperature rise. This was the case described in Section 4.1, and the temperature rise can be again obtained simply as a product of the (constant) power input and that of the overall thermal resistance, $\Delta T = PR_{th}$.

# 5. Energy Transport in Nanotubes, Graphene, and Nanowires

## 5.1 Carbon Nanotubes and Graphene

Carbon nanotubes (CNTs) and more recently graphene have emerged as new materials for electronics [101, 104, 105]. Graphene is a monolayer of graphite, a single sheet of hexagonally-arranged carbon atoms (Fig. 6a), first isolated five years ago [106]. Carbon nanotubes are cylindrical molecules with diameters ~1-3 nm, essentially "rolled-up" graphene obtained from chemical reactions [107]. Both materials have generated considerable interest because they demonstrate high mobility, thermal conductivity, and (particularly for graphene) potential for integration with planar CMOS. Their excellent electrical and thermal properties make them especially attractive for energy efficient nanoelectronics. From this point of view, compared to traditional silicon, devices based on carbon nanomaterials benefit from:

(1) Symmetric energy bands (Fig. 6b) and equal electron/hole mobility, both 10-100x higher than silicon [108, 109], indicating less scattering and lower power dissipation.

(2) Strong $sp^2$ bonds leading to thermal conductivity ~10-20x higher than silicon, 5-10x higher than copper, and comparable to that of isotopically pure diamond [80, 81, 110].

(3) Optical phonon energy 3x higher than in silicon (~180 meV vs. ~60 meV, Fig. 6c), suggesting they are less occupied and less likely to scatter electrons at low fields and near room temperature [30].

Energy dissipation in CNTs and graphene occurs when electric fields accelerate charge carriers (electrons or holes), which then scatter with lattice phonons. It has been generally assumed that the strongest high-field relaxation mechanism is that with optical phonons (OP), which have high energy $\hbar\omega_{OP} \approx$ 160–200 meV [111-115]. For instance, heating and scattering with hot OPs is responsible for negative differential conductance observed in suspended CNTs [36, 116-118], as shown in Fig. 7d. For substrate-supported CNTs, theoretical work has suggested that energy dissipation could also occur directly with "remote" polar surface-optical (SO) phonon modes in the substrate, bypassing heating of the CNT [119, 120] (Fig. 7b). For example, the lowest SO phonons of several substrates have energies of approximately $\hbar\omega_{SO} \approx$ 48 meV for $Al_2O_3$, 56 meV for $SiO_2$, and 81 meV for AlN [55]. These are significantly lower than the CNT optical phonons, and may allow the device to directly dissipate part of its energy into the nearby insulator. A similar result has also been predicted for graphene on insulators like $SiO_2$ or $HfO_2$ [121], and has been well-known in silicon inversion layers with high-K dielectrics [55, 122]. However, evidence of direct energy dissipation with substrate phonons is presently lacking, perhaps in large part due to the expected exponential dependence of electron-SO coupling on the CNT-substrate distance [120]. Thus, surface roughness, which for CNTs and graphene can be of the order of the device "thickness," or adsorbed water and impurities could significantly mask the electron-SO coupling. Nevertheless, a few studies do reveal phonon non-equilibrium in CNTs [91] and a reduction of mobility in graphene [109, 123] which have been attributed to scattering with substrate SO phonons. Interestingly, although not yet confirmed, a trade-off appears to exist between the SO scattering effect on mobility and power dissipation. Soft, low-energy SO pho-nons would limit mobility, but at the same time enhance direct CNT cooling with the substrate.

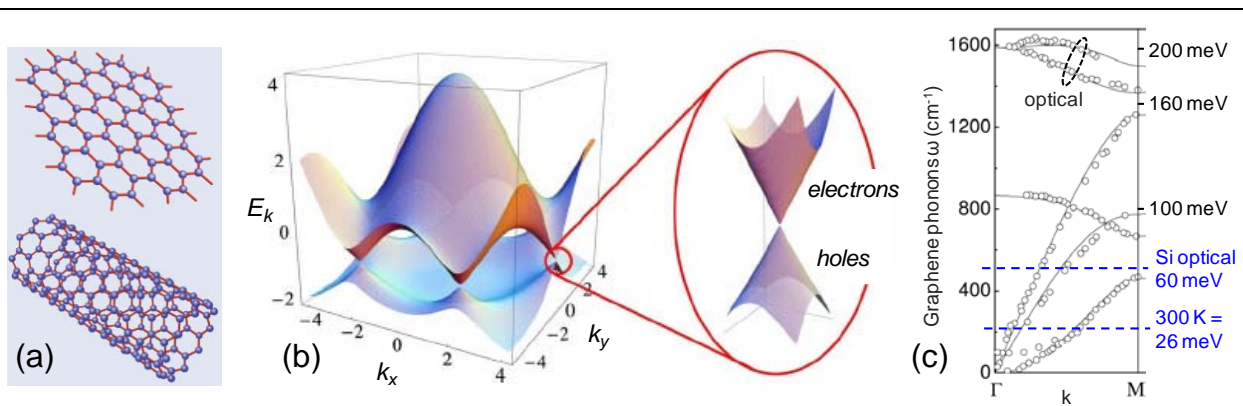

**Fig. 6.** (a) Lattice structure of graphene (top) and carbon nanotubes (bottom) [101]. (b) Energy dispersion in graphene. Right: energy bands close to Dirac point show linear dispersion [102]. (c) Phonon dispersion in graphene [103], showing optical phonon energy much greater than in silicon.

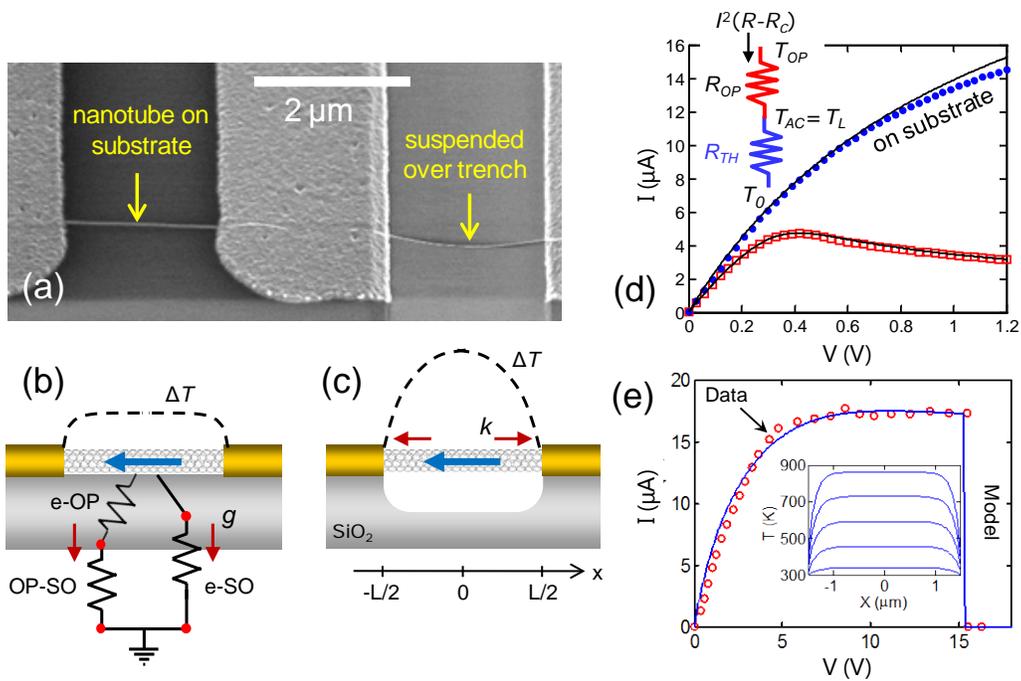

**Fig. 7.** Energy dissipation in substrate-supported and freely suspended metallic CNTs. (a) SEM image of a single CNT that is part suspended and part supported [36]. Schematic energy dissipation in (b) supported and (c) suspended CNTs. Red arrows show direction of heat flow. In supported CNT electrons scatter with device OPs, which then couple with the substrate. Electrons may also dissipate energy directly to SOs of the substrate [120]. (d) Measured data (symbols) and model (lines) of the CNT segments in (a). (e) Excellent agreement between data and model up to breakdown for another substrate-supported CNT (L = 3 μm) [30]. Inset shows temperature profile along CNT at applied voltages from 3–15 V, in 3 V increments.

More can be understood about energy dissipation and transport in 1-dimensional CNT and graphene devices by measuring their temperature during Joule heating by electrical current flow. Assuming a uniform heating rate per unit length $P/L \approx I^2(R-R_C)/L$ where $R$ is the total electrical resistance and $R_C$ is the combined electrical resistance of the two contacts, the temperature profile $T(x)$ of a substrate-supported metallic CNT (Fig. 7b) can be analytically obtained as [30, 87]:

$$T(x) = T_0 + \frac{P}{gL}\left[1 - \frac{\cosh(x/L_H)}{\cosh(L/2L_H)}\right] \tag{11}$$

for $-L/2 < x < L/2$, where $T_0$ is the temperature of the contacts at the two ends, $L_H = (kA/g)^{1/2} \approx 0.2$ μm is the characteristic thermal length along the CNT [124], $k$ is the thermal conductivity, and $g$ is the net heat dissipated to the substrate per unit length. In practice, a direct measurement of this temperature profile is very challenging due to the small CNT diameter. However, a scanning thermal microscopy technique (SThM) was recently demonstrated [87], showing agreement with the "flattened" temperature profile sketched in Fig. 7b and obtained when the CNT length is much longer than the characteristic thermal length ($L \gg L_H$). The drawbacks of SThM on CNTs are a relatively large uncertainty in its calibration and in the temperature values obtained.

Another measurement technique that can be applied to suspended CNTs uses the temperature dependence of the Raman-active phonon G-bands [91, 125, 126]. The temperature profile of the freely suspended metallic CNT (Fig. 7c) is a simple inverted parabola expressed as [36]:

$$T(x) = T_0 + \frac{P}{2kAL}\left[\left(\frac{L}{2}\right)^2 - x^2\right] \tag{12}$$

where again $-L/2 < x < L/2$ and $T_0$ is the temperature of the contacts at the two ends, and the power dissipation is assumed uniform along the CNT length. Analytic expressions for this temperature profile can also be obtained for a few cases of temperature-dependent thermal conductivity $k$ [127]. The Raman-thermometry technique has directly measured this steeply varying temperature profile in suspended CNTs [125], and has confirmed the phonon non-equilibrium between optical and acoustic modes previously suggested on the basis of electrical characteristics alone (Fig. 7d) [36]. In the same study [125], the authors also provided direct measurement of uneven thermal contact resistance between CNTs and their metal electrodes, with $R_{th,L} \approx 8 \times 10^7$ K/W (left electrode) and $R_{th,R} \approx 10^7$ K/W (right electrode) for the particular CNT under study. These are consistent with other estimates based on SThM (1.4–13.8 $\times 10^7$ K/W) [87] and breakdown thermometry (~1.2 $\times 10^7$ K/W) [86], but provide insight into the uneven quality of the two thermal contacts. The drawbacks of the Raman thermometry approach are its applicability to only those few CNTs resonant with the laser energy, and the relatively larger lateral resolution (~0.5 μm) compared to the SThM approach.

A third approach for substrate-supported CNTs can be referred to as "breakdown thermometry" and relies on measuring the CNT current at high voltage, including the Joule breakdown power. The voltage is gradually raised until the CNT breaks (Fig. 7e), while its *I-V* curve and breakdown power are compared to a transport model which includes heat dissipation. This was the approach adopted by Maune [78] and Pop [30, 86]. The breakdown temperature of CNTs is known to be approximately $T_{BD} \approx 600\ ^oC$ in air [128] and correlating it with the dissipated power yields the thermal coupling between CNT and substrate. This quantity is often given as a thermal conductance per CNT length, $g$ in $Wm^{-1}K^{-1}$ (Eq. 11 above). The three approaches have yielded somewhat different values of CNT-SiO$_2$ thermal coupling, with SThM suggesting $g \approx 0.007$-$0.06$ $Wm^{-1}K^{-1}$ [87], Raman thermometry yielding 0.03-0.1 $Wm^{-1}K^{-1}$ [91], and breakdown studies giving 0.1-0.2 $Wm^{-1}K^{-1}$ [78, 86]. This discrepancy is not yet resolved, because the electronic contribution to thermal coupling, if any, has not yet been isolated (Fig. 7b), and the various measurements were made on CNTs with different diameter, substrate surface roughness and ambient temperatures. For instance, breakdown thermometry yields the thermal coupling at the elevated breakdown temperatures, closer to 600 $^oC$ (although the average CNT temperature is lower). Moreover, it is relevant to point out that such nanoscale thermal coupling will be strongly dependent on the CNT diameter and substrate surface roughness of the measured samples, as recently revealed by molecular dynamics (MD) simulations [129]. The dissipation will also depend on the substrate material itself (e.g. quartz vs. SiO$_2$) as the phonon dispersion, surface optical phonons, and interface bonding between CNTs and substrate are expected to change. Such studies remain to be investigated.

### *5.2 Nanowires*

Unlike carbon nanotubes, nanowires are not truly 1-dimensional structures. Rather, they can be thought of as shrunken 3-dimensional rods which exhibit 1-dimensional features (e.g. energy sub-bands) when their diameters are brought to sizes comparable to the electron or phonon wavelength (~5-10 nm). In addition, nanowire surfaces are typically very good electron and phonon scatterers, such that nanowire mobility and thermal conductivity are universally lower than the bulk mobility or thermal conductivity of the same material (Fig. 5c-d and Fig. 8). For instance, the electrical resistivity of Cu interconnects with line widths below 100 nm increases by more than a factor of two from the bulk Cu resistivity [130], and the resistivity for line widths near 40 nm was shown to increase by at least a factor of four [131, 132] at room temperature.

The thermal conductivity of nanowires is similarly influenced by their surface. The thermal conductivity of bulk crystalline silicon is nearly 150 $Wm^{-1}K^{-1}$ at room temperature, which is reduced by one order of magnitude in nanowires with diameter below $D \approx 35$ nm (Fig. 5d and Fig. 8a) [82]. In simple terms this can be understood if we express the thermal conductivity as

$$k = \frac{1}{3}Cv\lambda \simeq \frac{1}{3}\int_i C_\omega v_\omega \left( \frac{A_1}{D} + \frac{A_2}{\lambda_0} + A_3 N_i \right)^{-1} d\omega \qquad (13)$$

where $C_\omega$ is the heat capacity per unit frequency mode, $v_\omega$ is the mode velocity, $\lambda_0$ is the phonon mean free path (MFP) in the bulk material, $N_i$ is the impurity concentration, and $A_{1-3}$ are fitting parameters [133-136]. The term in parenthesis represents the various contributions to the phonon MFP, including classical boundary scattering which becomes the limiting term when the diameter $D$ falls below the range of the bulk phonon MFP, $\lambda_0 \approx 0.1$–1 μm [54]. Variations on the simple expression above have generally been successful in reproducing experimental

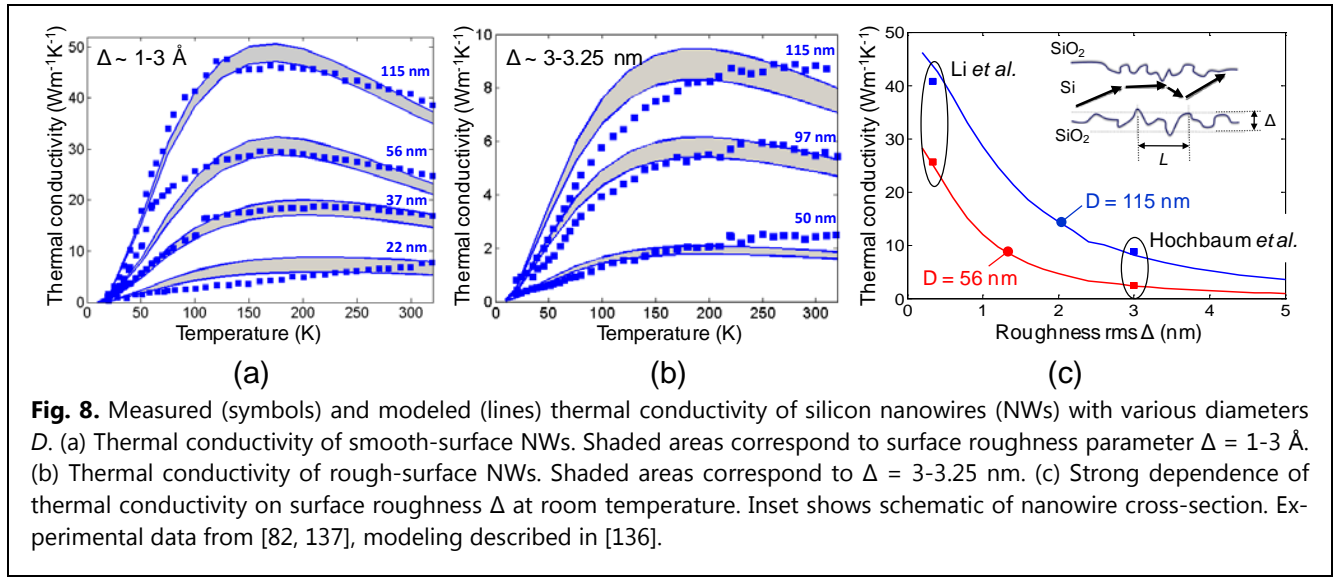

**Fig. 8.** Measured (symbols) and modeled (lines) thermal conductivity of silicon nanowires (NWs) with various diameters $D$. (a) Thermal conductivity of smooth-surface NWs. Shaded areas correspond to surface roughness parameter $\Delta$ = 1–3 Å. (b) Thermal conductivity of rough-surface NWs. Shaded areas correspond to $\Delta$ = 3–3.25 nm. (c) Strong dependence of thermal conductivity on surface roughness $\Delta$ at room temperature. Inset shows schematic of nanowire cross-section. Experimental data from [82, 137], modeling described in [136].

data on thermal conductivity of silicon nanowires [82] and thin films [83].

A new twist on our understanding of thermal transport in surface-limited nanowires was presented by the recent results of Chen, Hochbaum [137, 138] and Boukai [139], who found that silicon nanowires of diameter $D$ < 50 nm with very rough surfaces have thermal conductivity *one hundred* times lower than that of bulk crystalline silicon (Fig. 8b). This approaches the thermal conductivity of amorphous silicon, and cannot be understood on the basis of classical boundary scattering alone (the $1/D$ term in Eq. 13). To explain it, Martin et al [136] proposed an additional scattering term which relies on a description of the surface through a roughness height $\Delta$ and autocovariance length $L$ (Fig. 8c). This accounts for the fact that phonons "see" the rough nanowire as a series of constrictions along their propagation direction. Unlike classical boundary scattering, the roughness scattering rate is strongly frequency dependent ($\sim\omega^2$), with little impact on long-wavelength phonons. Using a full phonon dispersion for silicon, Martin et al [136] found an approximately quadratic dependence of nanowire thermal conductivity on diameter and roughness as $(D/\Delta)^2$. While the exact surface roughness in the experiments is difficult to measure, it appears that the augmented model does predict a very strong role for it, particularly for smaller diameter nanowires (Fig. 8b,c). If the electrical conductivity can be preserved, arrays of such nanostructures could form efficient thermoelectric devices, a domain of ongoing investigations.

## 6. Thermal Rectification

In addition to impeding or enhancing electrical and thermal transport, nanoscale structuring could also be used to introduce thermal flow asymmetry. This is the thermal equivalent of the electrical p–n diode, a two-terminal device that achieves greater heat flux in one direction than another, e.g. $Q_{BA} > Q_{AB}$ for the same temperature difference $\Delta T$, where A and B are its two terminals (Fig. 9) [140]. Thus, we can define the rectification ratio $\gamma = Q_{BA}/Q_{AB} > 1$ which will be referred to below. In the context of electronics and systems, thermal rectifiers could enable processing thermal flow independently from electron current, or the possibility of phononic devices and thermal logic [141]. In the context of heat engines or buildings, such thermal rectifiers could act as energy harvesting materials, exploiting naturally occurring temperature gradients with the environment.

A certain amount of thermal rectification can be achieved between two bulk materials with strongly different thermal conductivity dependence on temperature [142]. This is in fact a classical Fourier law effect, first observed in the 1970s [143, 144] and recently reexamined by Dames [142] and Kobayashi et al [145]. As shown in Fig. 9a, the two sides of the junction require two materials with strongly decreasing (A) and increasing (B) temperature dependence of thermal conductivity. When B is the "hot" and A is the "cold" terminal ($\Delta T = T_B - T_A$), both materials are in a region of high thermal conductivity, and heat flows more easily from B to A (top, wider arrow in Fig. 9a). When the terminals are reversed ($\Delta T = T_A - T_B$), both materials are in a region of lower thermal conductivity, and heat flow is impeded between A and B (bottom, thinner arrow). This leads to an asymmetric heat flux, $Q_{BA} > Q_{AB}$ for the same temperature difference $\Delta T$. Such thermal rectification was recently ob-served by Kobayashi et al [145], who noted a heat flux ratio $\gamma \approx 1.43$ between $LaCoO_3$ and $La_{0.7}Sr_{0.3}CoO_3$ in the temperature range 70–100 K.

Suggestions of nanostructures for thermal rectification have only appeared in the past few years. Several theoretical proposals [146, 149] have been put forward, estimating thermal rectification as high as $\gamma \approx$ 100-1000 between two directions. Thermal rectification could be achieved through nanoscale structuring, e.g. inducing nano-indentations that preferentially scatter phonons [147], through geometrical heat "funneling" e.g. in the case of a carbon nanocone [146], or by asymmetric mass-loading [148], as illustrated in Fig. 9b-d. Most suggestions based on nanoscale thermal rectification have been purely theoretical, requiring engineering of coupling and layout between individual atoms, which are very difficult to achieve in practice. The only nanoscale thermal rectifier demonstrated to date at room temperature [148] has shown thermal rectification ratio $\gamma \approx 1.02$–$1.07$. This device relied on suspended, several microns-long multi-wall carbon or boron nitride nanotubes that were unevenly mass-loaded with a much heavier molecular species, $C_9H_{16}Pt$ (Fig. 9d). In this case, heat flow preference was from the high- to the low-mass regions of the loaded nanotube. Thermal rectification at very low-temperature (80 mK) and high magnetic field (10 T) has also been demonstrated in single quantum dots, with heat flow asymmetry $\gamma \approx 1.11$ [150]. Clearly, much work remains to be done to achieve a practical, room-temperature thermal diode with rectification up to theoretically-predicted ($\gamma \approx$ 100-1000) limits.

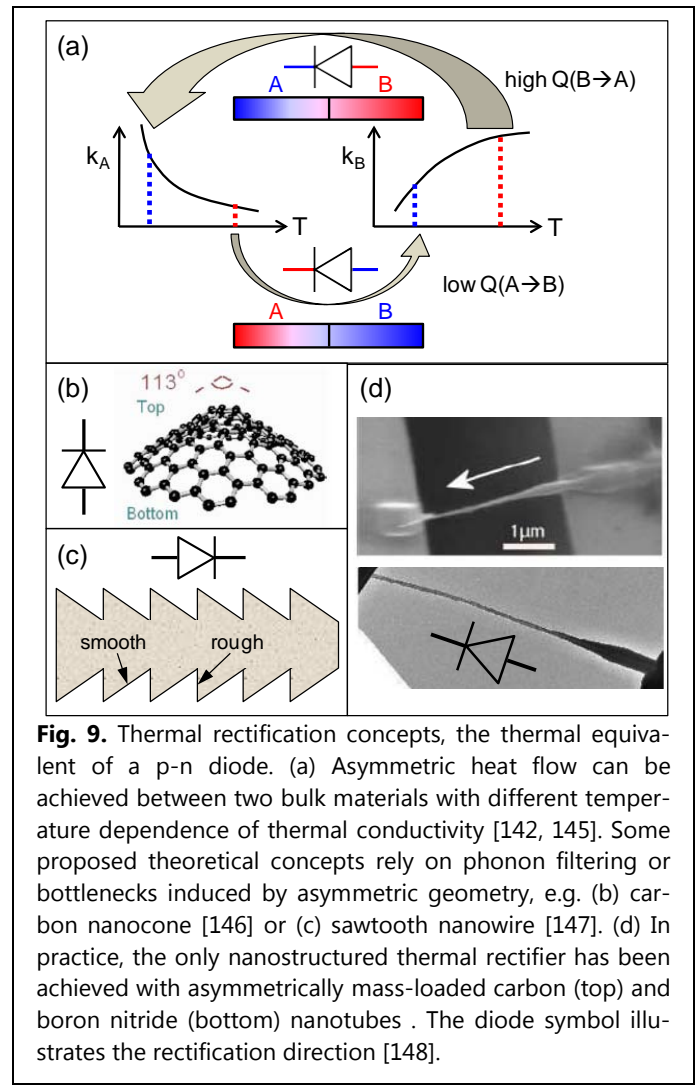

**Fig. 9.** Thermal rectification concepts, the thermal equivalent of a p-n diode. (a) Asymmetric heat flow can be achieved between two bulk materials with different temperature dependence of thermal conductivity [142, 145]. Some proposed theoretical concepts rely on phonon filtering or bottlenecks induced by asymmetric geometry, e.g. (b) carbon nanocone [146] or (c) sawtooth nanowire [147]. (d) In practice, the only nanostructured thermal rectifier has been achieved with asymmetrically mass-loaded carbon (top) and boron nitride (bottom) nanotubes . The diode symbol illustrates the rectification direction [148].

## **7. Interfaces**

Nanoscale 1- and 2-dimensional devices have very little "bulk" and thus their behavior is strongly dominated by their interfaces in terms of energy transport and dissipation. This is particularly evident for thermal conductivity in nanowires, or thermal dissipation in nanotubes, as outlined above. While the experimental data and theoretical understanding of nanoscale device interfaces are just beginning to take shape, a wealth of experimental data exists for energy transport across interfaces between otherwise bulk solids. This is summarized in Fig. 10. The interface thermal conductance relates the heat flux $Q$ crossing an interface to the temperature drop $\Delta T$ at the interface, $Q = G\Delta T$. Thermal interfaces are most often studied with pump-probe optical techniques such as time-domain thermoreflectance (TDTR), picosecond transient absorption [151-156], or the 3-omega method [157-161]. As summarized in Fig. 10, these studies have shown that the thermal conductance of bulk solid interfaces at room temperature spans a relatively limited range, but depends on composition of the interface at the level of a single molecular layer [162].

The range of available thermal conductances at material interfaces appears to approach two limits. At the upper end, the best measured thermal interfaces are between metals with good intrinsic thermal properties, e.g. Al/Cu. Thermal transport across these interfaces appears to be mediated by electrons, and the phonon contribution is less than 10% [163]. The best metal-dielectric thermal interfaces occur between materials with similar Debye temperatures $\Theta_D$ (such as highly annealed, epitaxial TiN on single crystal oxides [152]) and approach the limit of a phonon diffuse mismatch model (DMM) [164]. At the low end, a large Debye temperature mismatch between the two materials is expected to lead to significant thermal impedance, such as between Pb or Bi ($\Theta_D \sim 110$ K) and diamond ($\Theta_D \sim 2200$ K) [153]. The Debye temperature is a typical figure of merit for the difference in vibrational spectrum and the phonon density of states between the two sides of the interface. With regards to Fig. 10 it is interesting to note the physical meaning of the thermal interface conductance by relating it to the equivalent thermal conductance of a dielectric layer with thickness $d$ and thermal conductivity $k$, i.e. $G = k/d$. Thus, an interface with $G = 10$ MWm$^{-2}$K$^{-1}$ is equivalent to the thermal impedance of a 140 nm film of SiO$_2$ ($k_{ox} = 1.4$ Wm$^{-1}$K$^{-1}$) and $G = 100$ MWm$^{-2}$K$^{-1}$ is equivalent to the thermal impedance of 14 nm of SiO$_2$. This simple comparison also highlights the importance of such material interfaces in all nanometer scale devices and structures.

Among 1- or 2-dimensional conductors, some of the available data on interface thermal resistance for carbon nanotubes was already presented in Section 5.1, and data for the graphene/SiO$_2$ thermal interface is shown in Fig. 10. In particular, given comparable Debye temperature of CNTs and graphene with diamond, one expects a significant vibrational mismatch between them and other materials with typically lower $\Theta_D$. To compare the data available for CNTs with that of other material interfaces, the range of thermal conductance per unit length $g \sim 0.007$–$0.2$ Wm$^{-1}$K$^{-1}$ [78, 86, 87, 91] can be converted to $G \sim 4$–$100$ MWm$^{-2}$K$^{-1}$ per unit area when normalized by a 2 nm CNT diameter, well within the range of the experimental data in Fig. 10. In particular, given the large intrinsic thermal conductivity of CNTs it appears that their thermal resistance is nearly *always* dominated by their interface with the environment, both for solid [155] as well as liquid composites [154]. Nevertheless, little is yet known about how to engineer 1-dimensional interfaces between dissimilar materials such as CNTs and their dielectric environments, and much work remains to be done in this area.

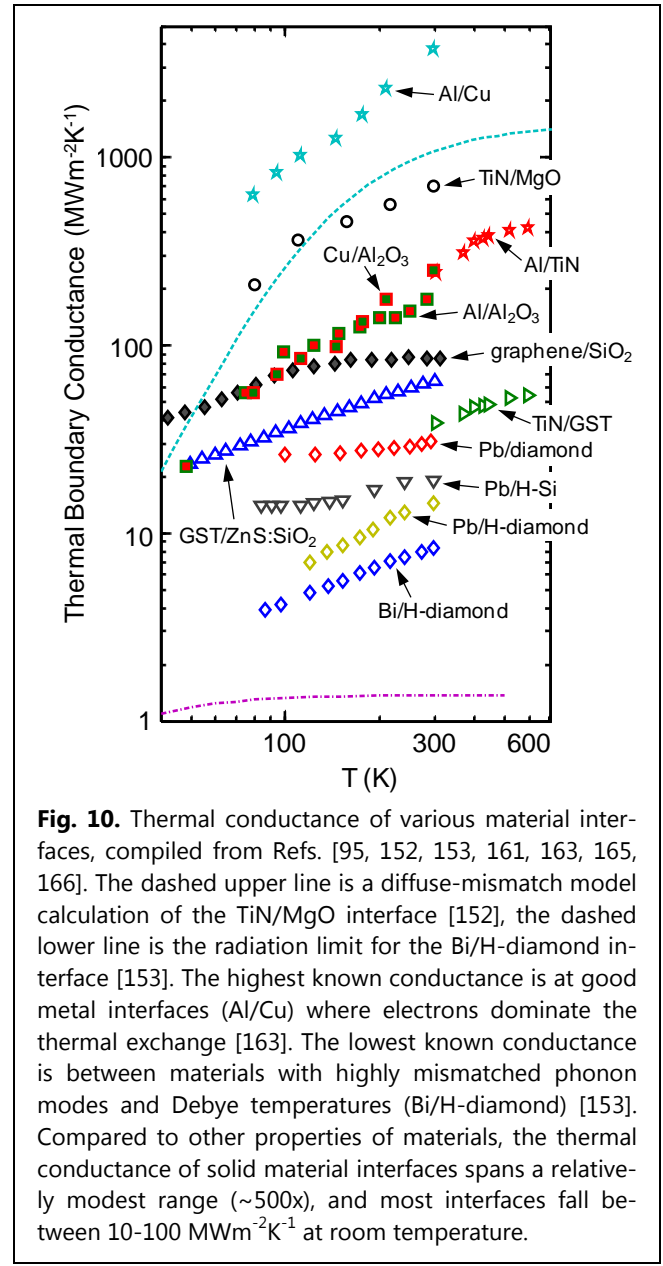

**Fig. 10.** Thermal conductance of various material interfaces, compiled from Refs. [95, 152, 153, 161, 163, 165, 166]. The dashed upper line is a diffuse-mismatch model calculation of the TiN/MgO interface [152], the dashed lower line is the radiation limit for the Bi/H-diamond interface [153]. The highest known conductance is at good metal interfaces (Al/Cu) where electrons dominate the thermal exchange [163]. The lowest known conductance is between materials with highly mismatched phonon modes and Debye temperatures (Bi/H-diamond) [153]. Compared to other properties of materials, the thermal conductance of solid material interfaces spans a relatively modest range (~500x), and most interfaces fall between 10–100 MWm$^{-2}$K$^{-1}$ at room temperature.

## 8. Conclusions

Understanding and controlling energy dissipation and transport in nanostructures will continue to be an area of rapid improvement and discovery, with applications ranging from low-power electronics, to energy-efficient data centers, and thermal energy harvesting. For instance, reducing power dissipation in electronics will impact a wide range of applications from mobile devices ($10^{-3}$ W) to massive data centers ($10^9$ W). Such developments would also improve weight payloads for mobile electronics and medical instrumentation (fewer heavy batteries), and potentially exploit ubiquitous thermal gradients –from sources as varied as car engines, power plants, or even body-environment temperature differences– as supplies of electrical energy.

On a large scale, by simple estimates even a 2× more energy-efficient transistor could lower nation-wide power use by over $10^{10}$ W if implemented today, which is a significant percentage of our national power budget. Given that energy use by electronics is on a trend that could reach 30% of the national electricity consumption by 2025, such progress is crucial to maintaining progress in a post-CMOS world, and has great environmental implications as well. In addition, just over half the world energy is wasted as heat (~$10^{13}$ W), from nuclear power plants and factories, to car engines and the power bricks on our laptops. Efficiently reclaiming even a small percentage of such wasted heat would itself nearly satisfy the electricity needs of our planet.

## 9. Acknowledgement

I am indebted to Profs. D. Cahill, H. Dai, C. Dames, D. Jena, W. King, J.-P. Leburton, J. Lyding and U. Ravaioli for many valuable discussions. I also thank Dr. M.-H. Bae for feedback on an earlier manuscript draft. This review was in part supported by the Nanoelectronics Research Initiative (NRI), the DARPA Young Faculty Award No. HR0011-08-1-0035, the Office of Naval Research grant No. N00014-09-1-0180, the National Science Foundation grant CCF 08-29907, Intel Corp. and Northrop Grumman.

## References

1. EPA Report on Server and Data Center Energy Efficiency, *http://www.energystar.gov/index.cfm?c=prod_development.server_efficiency_study*

2. Pop, E., Sinha, S. & Goodson, K.E. Heat generation and transport in nanometer-scale transistors. *Proc. IEEE* **94**, 1587-1601 (2006).

3. Cavin, R., Zhirnov, V., Herr, D., Avila, A. & Hutchby, J. Research directions and challenges in nanoelectronics. *Journal of Nanoparticle Research* **8**, 841-858 (2006).

4. Haensch, W. et al. Silicon CMOS devices beyond scaling. *IBM J. Res. Dev.* **50**, 339 (2006).

5. Intel products, *http://ark.intel.com/*

6. PC Energy Report 2009, *http://www.climatesaverscomputing.org*

7. Energy & Environment Department at LBNL, *http://eed.llnl.gov/flow/02flow.php*

8. The Carbon Footprint of Email Spam Report, *http://www.mcafee.com*

9. Roy, K. & Prasad, S. Low-power CMOS VLSI circuit design. (Wiley, 2000).

10. Hanson, S. et al. Ultralow-Voltage Minimum-Energy CMOS. *IBM J. Res. Dev.* **50**, 469-490 (2006).

11. Zhai, B., Blaauw, D., Sylvester, D. & Flautner, K. The limit of dynamic voltage scaling and insomniac dynamic voltage scaling. *Very Large Scale Integration (VLSI) Systems, IEEE Transactions on* **13**, 1239-1252 (2005).

12. Akinwande, D., Liang, J., Chong, S., Nishi, Y. & Wong, H.S.P. Analytical ballistic theory of carbon nanotube transistors: Experimental validation, device physics, parameter extraction, and performance projection. *Journal of Applied Physics* **104**, 124514-124517 (2008).

13. Jenkins, K.A. & Rim, K. Measurement of the effect of self-heating in strained-silicon MOSFETs. *IEEE Electron Device Letters* **23**, 360-362 (2002).

14. Tenbroek, B., Lee, M.S.L., Redman-White, W., Bunyan, R.J.T. & Uren, M.J. Self-heating effects in SOI MOSFETs and their measurement by small signal conductance techniques. *IEEE Trans. Electron Devices* **43**, 2240-2248 (1996).

15. Clemente, S. Transient thermal response of power semiconductors to short power pulses. *Power Electronics, IEEE Transactions on* **8**, 337-341 (1993).

16. Yan, H. et al. Time-resolved Raman spectroscopy of optical phonons in graphite: Phonon anharmonic coupling and anomalous stiffening. *Physical Review B* **80**, 121403-121404 (2009).

17. Menéndez, J. & Cardona, M. Temperature dependence of the first-order Raman scattering by phonons in Si, Ge, and alpha-Sn: Anharmonic effects. *Physical Review B* **29**, 2051 (1984).

18. Letcher, J.J., Kang, K., Cahill, D.G. & Dlott, D.D. Effects of high carrier densities on phonon and carrier lifetimes in Si by time-resolved anti-Stokes Raman scattering. *Applied Physics Letters* **90**, 252104-252103 (2007).

19. Sinha, S., Schelling, P.K., Phillpot, S.R. & Goodson, K.E. Scattering of g-process longitudinal optical phonons at hotspots in silicon. *Journal of Applied Physics* **97**, 023702-023709 (2005).

20. Ledgerwood, M.L. & van Driel, H.M. Picosecond phonon dynamics and self-energy effects in highly photoexcited germanium. *Physical Review B* **54**, 4926 (1996).

21. Ioffe Institute, *http://www.ioffe.rssi.ru/SVA/NSM/Semicond/index.html*

22. Koswatta, S.O., Lundstrom, M.S. & Nikonov, D.E. Band-to-Band Tunneling in a Carbon Nanotube Metal-Oxide-Semiconductor Field-Effect Transistor Is Dominated by Phonon-Assisted Tunneling. *Nano Letters* **7**, 1160-1164 (2007).

23. Buttiker, M. Role of quantum coherence in series resistors. *Phys. Rev. B* **33**, 3020 (1986).

24. Das, M.P. & Green, F. Landauer formula without Landauer's assumptions. *J. Phys.: Cond. Matt.* **15**, 687-693 (2003).

25. Datta, S. Electronic Transport in Mesoscopic Systems. (Univ. Press, Cambridge; 1995).

26. Datta, S. Quantum Transport: Atom to Transistor. (Cambridge Univ. Press, 2006).

27. Lake, R. & Datta, S. Energy balance and heat exchange in mesoscopic systems. *Phys. Rev. B* **46**, 4757 (1992).

28. Ouyang, Y. & Guo, J. Heat dissipation in carbon nanotube transistors. *Appl. Phys. Lett.* **89**, 183122-183124 (2006).

29. Park, J.Y. et al. Electron-phonon scattering in metallic single-walled carbon nanotubes. *Nano Letters* **4**, 517 (2004).

30. Pop, E., Mann, D., Goodson, K. & Dai, H. Electrical and thermal transport in metallic single-wall carbon nanotubes on insulating substrates. *J. Appl. Phys.* **101**, 093710 (2007).
31. Liao, A., Zhao, Y. & Pop, E. Avalanche-Induced Current Enhancement in Semiconducting Carbon Nanotubes. *Physical Review Letters* **101**, 256804-256804 (2008).
32. Chen, Y.-C., Zwolak, M. & Di Ventra, M. Local Heating in Nanoscale Conductors. *Nano Letters* **3**, 1691-1694 (2003).
33. Segal, D. & Nitzan, A. Heating in current carrying molecular junctions. *The Journal of Chemical Physics* **117**, 3915-3927 (2002).
34. Galperin, M., Saito, K., Balatsky, A.V. & Nitzan, A. Cooling mechanisms in molecular conduction junctions. *Physical Review B (Condensed Matter and Materials Physics)* **80**, 115427-115412 (2009).
35. Galperin, M., Ratner, M.A. & Nitzan, A. Molecular transport junctions: vibrational effects. *Journal of Physics: Condensed Matter*, 103201 (2007).
36. Pop, E. et al. Negative differential conductance and hot phonons in suspended nanotube molecular wires. *Phys. Rev. Lett.* **95**, 155505 (2005).
37. D'Agosta, R., Sai, N. & Di Ventra, M. Local Electron Heating in Nanoscale Conductors. *Nano Letters* **6**, 2935-2938 (2006).
38. Koswatta, S.O., Lundstrom, M.S. & Nikonov, D.E. Influence of phonon scattering on the performance of p-i-n band-to-band tunneling transistors. *Applied Physics Letters* **92**, 043125-043123 (2008).
39. Pop, E., Rowlette, J., Dutton, R.W. & Goodson, K.E. in Intl. Conf. on Simulation of Semiconductor Processes and Devices (SISPAD) 307-310Tokyo, Japan; 2005).
40. Stettler, M.A., Alam, M.A. & Lundstrom, M.S. A critical examination of the assumptions underlying macroscopic transport equations for silicon devices. *Electron Devices, IEEE Transactions on* **40**, 733-740 (1993).
41. Vashaee, D. & Shakouri, A. Improved Thermoelectric Power Factor in Metal-Based Superlattices. *Physical Review Letters* **92**, 106103 (2004).
42. Mahan, G.D. & Woods, L.M. Multilayer Thermionic Refrigeration. *Physical Review Letters* **80**, 4016 (1998).
43. Pipe, K.P., Ram, R.J. & Shakouri, A. Internal cooling in a semiconductor laser diode. *IEEE Phot. Tech. Lett.* **14**, 453-455 (2002).
44. Shakouri, A. & Bowers, J.E. Heterostructure integrated thermionic coolers. *Applied Physics Letters* **71**, 1234-1236 (1997).
45. Shakouri, A., Lee, E.Y., Smith, D.L., Narayanamurti, V. & Bowers, J.E. Thermoelectric Effects in Submicron Heterostructure Barriers. *Microscale Thermophysical Engineering* **2**, 37-47 (1998).
46. Xu, X., Gabor, N.M., Alden, J.S., van der Zande, A.M. & McEuen, P.L. Photo-Thermoelectric Effect at a Graphene Interface Junction. *Nano Letters ASAP*, http://dx.doi.org/10.1021/nl903451y (2009).
47. Wachutka, G.K. Rigorous thermodynamic treatment of heat generation and conduction in semiconductor device modeling. *IEEE Transactions on Computer-Aided Design* **9**, 1141-1149 (1990).
48. Lindefelt, U. Heat generation in semiconductor devices. *J. Appl. Phys.* **75**, 942-957 (1994).
49. Sverdrup, P.G., Ju, Y.S. & Goodson, K.E. Sub-continuum simulations of heat conduction in silicon-on-insulator transistors. *J. Heat Transfer* **123**, 130-137 (2001).
50. Lai, J. & Majumdar, A. Concurrent thermal and electrical modeling of sub-micrometer silicon devices. *J. Appl. Phys.* **79**, 7353-7361 (1996).
51. Pop, E., Dutton, R.W. & Goodson, K.E. Analytic band Monte Carlo model for electron transport in Si including acoustic and optical phonon dispersion. *J. Appl. Phys.* **96**, 4998 (2004).
52. Ju, Y.S. & Goodson, K.E. Phonon scattering in silicon thin films with thickness of order 100 nm. *Appl. Phys. Lett.* **74**, 3005-3007 (1999).
53. Mazumder, S. & Majumdar, A. Monte Carlo study of phonon transport in solid thin films including dispersion and polarization. *J. Heat Transfer* **123**, 749-759 (2001).
54. Henry, A.S. & Chen, G. Spectral Phonon Transport Properties of Silicon Based on Molecular Dynamics Simulations and Lattice Dynamics. *Journal of Computational and Theoretical Nanoscience* **5**, 141-152 (2008).
55. Fischetti, M.V., Neumayer, D.A. & Cartier, E.A. Effective electron mobility in Si inversion layers in MOS systems with a high-K insulator: the role of remote phonon scattering. *J. Appl. Phys.* **90**, 4587 (2001).
56. Artaki, M. & Price, P.J. Hot phonon effects in silicon field-effect transistors. *J. Appl. Phys.* **65**, 1317-1320 (1989).
57. Lugli, P. & Goodnick, S.M. Nonequilibrium longitudinal-optical phonon effects in GaAs-AlGaAs quantum wells. *Physical Review Letters* **59**, 716 (1987).
58. Pop, E., Dutton, R.W. & Goodson, K.E. Monte Carlo simulation of Joule heating in bulk and strained silicon. *Appl. Phys. Lett.* **86**, 082101 (2005).
59. Jacoboni, C. & Reggiani, L. The Monte Carlo method for the solution of charge transport in semiconductors with applications to covalent materials. *Rev. Mod. Phys.* **55**, 645-705 (1983).
60. Fischetti, M.V. & Laux, S.E. Monte carlo analysis of electron transport in small semiconductor devices including band-structure and space-charge effects. *Physical Review B* **38**, 9721 (1988).
61. Rowlette, J.A. & Goodson, K.E. Fully Coupled Nonequilibrium Electron-Phonon Transport in Nanometer-Scale Silicon FETs. *Electron Devices, IEEE Transactions on* **55**, 220-232 (2008).
62. Sinha, S., Pop, E., Dutton, R.W. & Goodson, K.E. Non-equilibrium phonon distributions in sub-100 nm silicon transistors. *J. Heat Transfer* **128**, 638-647 (2006).

63. Raleva, K., Vasileska, D., Goodnick, S.M. & Nedjalkov, M. Modeling Thermal Effects in Nanodevices. *Electron Devices, IEEE Transactions on* **55**, 1306-1316 (2008).
64. Vasileska, D., Raleva, K. & Goodnick, S. Modeling heating effects in nanoscale devices: the present and the future. *Journal of Computational Electronics* **7**, 66-93 (2008).
65. Lundstrom, M. & Ren, Z. Essential physics of carrier transport in nanoscale MOSFETs. *IEEE Trans. Electron Devices* **49**, 133-141 (2002).
66. Wang, Y., Cheung, P., Oates, A. & Mason, P. in Reliability physics symposium, 2007. proceedings. 45th annual. ieee international 258-263 2007).
67. Bunyan, R.J.T., Uren, M.J., Alderman, J.C. & Eccleston, W. Use of noise thermometry to study the effects of self-heating in submicrometer SOI MOSFETs. *IEEE Electron Device Letters* **13**, 279 (1992).
68. Darwish, A.M., Bayba, A.J. & Hung, H.A. Accurate determination of thermal resistance of FETs. *IEEE Trans. Microwave Theory and Tech.* **53**, 306 (2005).
69. Jenkins, K.A. & Sun, J.Y.-C. Measurement of I-V curve of silicon-on-insulator (SOI) MOSFETs without self-heating. *IEEE Electron Device Letters* **16**, 145 (1995).
70. Jin, W., Liu, W., Fung, S.K.H., Chan, P.C.H. & Hu, C. SOI thermal impedance extraction methodology and its significance for circuit simulation. *IEEE Trans. Electron Devices* **48**, 730-736 (2001).
71. Joy, R.C. & Schlig, E.S. Thermal properties of very fast transistors. *IEEE Trans. Electron Devices* **17**, 386 (1970).
72. Lee, T.-Y. & Fox, R.M. in IEEE Intl. SOI Conference 78 1995).
73. Mautry, P.G. & Trager, J. in Intl. Conf. on Microelectronic Test Struct., Vol. 3 221 1990).
74. Reyboz, M., Daviot, R., Rozeau, O., Martin, P. & Paccaud, M. in IEEE Intl. SOI Conference 159 2004).
75. Rinaldi, N. On the modeling of the transient thermal behavior of semiconductor devices. *IEEE Trans. Electron Devices* **48**, 2796 (2001).
76. Su, L.T., Chung, J.E., A., A.D., Goodson, K.E. & Flik, M.I. Measurement and modeling of self-heating in SOI NMOSFETs. *IEEE Trans. Electron Devices* **41**, 69-75 (1994).
77. Pop, E. Self-Heating and Scaling of Thin-Body Transistors, *Ph.D. Thesis*. (Stanford Univ., Stanford, CA; 2005), Available *http://poplab.ece.illinois.edu*
78. Maune, H., Chiu, H.-Y. & Bockrath, M. Thermal resistance of the nanoscale constrictions between carbon nanotubes and solid substrates. *Appl. Phys. Lett.* **89**, 013109 (2006).
79. Kim, P., Shi, L., Majumdar, A. & McEuen, P.L. Mesoscopic thermal transport and energy dissipation in carbon nanotubes. *Physica B: Condensed Matter* **323**, 67-70 (2002).
80. Pop, E., Mann, D., Wang, Q., Goodson, K.E. & Dai, H.J. Thermal conductance of an individual single-wall carbon nanotube above room temperature. *Nano Letters* **6**, 96-100 (2006).
81. Yu, C., Shi, L., Yao, Z., Li, D. & Majumdar, A. Thermal conductance and thermopower of an individual single-wall carbon nanotube. *Nano Letters* **5**, 1842-1846 (2005).
82. Li, D. et al. Thermal conductivity of individual silicon nanowires. *Appl. Phys. Lett.* **83**, 2934-2936 (2003).
83. Liu, W. & Asheghi, M. Thermal conduction in ultrathin pure and doped single-crystal silicon layers at high temperature. *J. Appl. Phys.* **98**, 123523 (2005).
84. Rego, L.G.C. & Kirczenow, G. Quantized Thermal Conductance of Dielectric Quantum Wires. *Phys. Rev. Lett.* **81**, 232 (1998).
85. Schwab, K., Henriksen, E.A., WorlocK, J.M. & Roukes, M.L. Measurement of the quantum of thermal conductance. *Nature* **2000**, 974 (2000).
86. Pop, E. The role of electrical and thermal contact resistance for Joule breakdown of single-wall carbon nanotubes. *Nanotechnology* **19**, 295202 (2008).
87. Shi, L. et al. Thermal probing of energy dissipation in current-carrying carbon nanotubes. *Journal of Applied Physics* **105**, 104306-104305 (2009).
88. Yovanovich, M.M., Culham, J.R. & Teertstra, P. Analytical modeling of spreading resistance in flux tubes, half spaces, and compound disks. *IEEE Trans. Components, Packaging and Manuf. Tech.* **21**, 168 (1998).
89. Micro Heat Transfer Lab (U. Waterloo), *http://www.mhtl.uwaterloo.ca/RScalculators.html*
90. Wang, X. et al. Room-Temperature All-Semiconducting Sub-10-nm Graphene Nanoribbon Field-Effect Transistors. *Physical Review Letters* **100**, 206803-206804 (2008).
91. Steiner, M. et al. Phonon populations and electrical power dissipation in carbon nanotube transistors. *Nature Nanotechnology* **4**, 320-324 (2009).
92. Reifenberg, J.P., Kencke, D.L. & Goodson, K.E. The impact of thermal boundary resistance in phase-change memory devices. *IEEE Elec. Dev. Lett.* **29**, 1112-1114 (2008).
93. Chen, I.R. & Pop, E. Compact Thermal Model for Vertical Nanowire Phase-Change Memory Cells. *IEEE Transactions on Electron Devices* **56**, 1523-1528 (2009).
94. Freitag, M. et al. Energy Dissipation in Graphene Field-Effect Transistors. *Nano Letters* **9**, 1883-1888 (2009).
95. Chen, Z., Jang, W., Bao, W., Lau, C.N. & Dames, C. Thermal contact resistance between graphene and silicon dioxide. *Applied Physics Letters* **95**, 161910-161913 (2009).

96. Amerasekera, A. & Duvvury, C. ESD in Silicon Integrated Circuits, Edn. 2nd. (Wiley, 2002).
97. Jenkins, K.A. & Franch, R.L. in IEEE Intl. SOI Conference 161-1632003).
98. Banerjee, K., Amerasekera, A., Cheung, N. & Hu, C. High-current failure model for VLSI interconnects under short-pulse stress conditions. *IEEE Electron Device Letters* **18**, 405 (1997).
99. Dwyer, V.M., Franklin, A.J. & Campbell, D.S. Thermal failure in semiconductor devices. *Solid-State Electronics* **33**, 553 (1990).
100. Min, Y.J., Palisoc, A.L. & Lee, C.C. Transient thermal study of semiconductor devices. *IEEE Trans. Components, Hybrids, Manufacturing Technol.* **13**, 980 (1990).
101. Castro Neto, A., Guinea, F. & Miguel, N. Drawing conclusions from graphene. *Physics World* (2006).
102. Neto, A.H.C., Guinea, F., Peres, N.M.R., Novoselov, K.S. & Geim, A.K. The electronic properties of graphene. *Reviews of Modern Physics* **81**, 109-154 (2009).
103. Yanagisawa, H. et al. Analysis of phonons in graphene sheets by means of HREELS measurement and *ab initio* calculation. *Surface and Interface Analysis* **37**, 133-136 (2005).
104. Dresselhaus, M.S., Dresselhaus, G. & Avouris, P. Carbon nanotubes: synthesis, structure, properties and applications. (Springer, Berlin; 2001).
105. Geim, A.K. & Novoselov, K.S. The rise of graphene. *Nature Materials* **6**, 183-191 (2007).
106. Novoselov, K.S. et al. Electric field effect in atomically thin carbon films. *Science* **306**, 666 (2004).
107. Iijima, S. Helical microtubules of graphitic carbon. *Nature* **354**, 56-58 (1991).
108. Durkop, T., Getty, S.A., Cobas, E. & Fuhrer, M.S. Extraordinary mobility in semiconducting carbon nanotubes. *Nano Letters* **4**, 35-39 (2004).
109. Morozov, S.V. et al. Giant Intrinsic Carrier Mobilities in Graphene and Its Bilayer. *Phys. Rev. Lett.* **100**, 016602 (2008).
110. Nika, D.L., Pokatilov, E.P., Askerov, A.S. & Balandin, A.A. Phonon thermal conduction in graphene: Role of Umklapp and edge roughness scattering. *Phys. Rev. B* **79**, 155413 (2009).
111. Yao, Z., Kane, C.L. & Dekker, C. High-field electrical transport in single-wall carbon nanotubes. *Physical Review Letters* **84**, 2941 (2000).
112. Javey, A. et al. High-field quasiballistic transport in short carbon nanotubes. *Physical Review Letters* **92**, 106804 (2004).
113. Kuroda, M.A., Cangellaris, A. & Leburton, J.-P. Nonlinear transport and heat dissipation in metallic carbon nanotubes. *Phys. Rev. Lett.* **95**, 266803 (2005).
114. Hasan, S., Alam, M.A. & Lundstrom, M.S. Simulation of carbon nanotube FETs including hot-phonon and self-heating effects. *IEEE Trans. Elec. Dev.* **54**, 2352 (2007).
115. Perebeinos, V., Tersoff, J. & Avouris, P. Electron-Phonon Interaction and Transport in Semiconducting Carbon Nanotubes. *Physical Review Letters* **94**, 086802 (2005).
116. Wang, X. et al. Electrically driven light emission from hot single-walled carbon nanotubes at various temperatures and ambient pressures. *Applied Physics Letters* **91**, 261102-261103 (2007).
117. Mann, D. et al. Electrically driven thermal light emission from individual single-walled carbon nanotubes. *Nat Nano* **2**, 33-38 (2007).
118. Mann, D., Pop, E., Cao, J., Wang, Q. & Goodson, K. Thermally and Molecularly Stimulated Relaxation of Hot Phonons in Suspended Carbon Nanotubes. *The Journal of Physical Chemistry B* **110**, 1502-1505 (2006).
119. Petrov, A.G. & Rotkin, S.V. Energy relaxation of hot carriers in single-wall carbon nanotubes by surface optical phonons of the substrate. *JETP Letters* **84**, 156-160 (2006).
120. Rotkin, S.V., Perebeinos, V., Petrov, A.G. & Avouris, P. An essential mechanism of heat dissipation in carbon nanotube electronics. *Nano Letters* **9**, 1850-1855 (2009).
121. Fratini, S. & Guinea, F. Substrate-limited electron dynamics in graphene. *Phys. Rev. B* **77**, 195415 (2008).
122. Chau, R. et al. High-K/metal-gate stack and its MOSFET characteristics. *Electron Device Letters, IEEE* **25**, 408-410 (2004).
123. Chen, J.-H., Jang, C., Xiao, S., Ishigami, M. & Fuhrer, M.S. Intrinsic and extrinsic performance limits of graphene devices on SiO2. *Nat Nano* **3**, 206-209 (2008).
124. Xiong, F., Liao, A. & Pop, E. Inducing Chalcogenide Phase Change with Ultra-Narrow Carbon Nanotube Heaters. *Appl. Phys. Lett.* **95**, 243103 (2009).
125. Deshpande, V.V., Hsieh, S., Bushmaker, A.W., Bockrath, M. & Cronin, S.B. Spatially Resolved Temperature Measurements of Electrically Heated Carbon Nanotubes. *Physical Review Letters* **102**, 105501-105504 (2009).
126. Hsu, I.K. et al. Optical measurement of thermal transport in suspended carbon nanotubes. *Applied Physics Letters* **92**, 063119-063113 (2008).
127. Huang, X.Y., Zhang, Z.Y., Liu, Y. & Peng, L.M. Analytical analysis of heat conduction in a suspended one-dimensional object. *Applied Physics Letters* **95**, 143109-143103 (2009).
128. Hata, K. et al. Water-assisted highly efficient synthesis of impurity-free single-walled carbon nanotubes. *Science* **306**, 1362-1364 (2004).
129. Ong, Z.-Y. & Pop, E. Molecular Dynamics Simulation of Thermal Boundary Conductance Between Carbon Nanotubes and SiO$_2$. *http://arxiv.org/abs/0910.2747* (2009).
130. Plombon, J.J., Andideh, E., Dubin, V.M. & Maiz, J. Influence of phonon, geometry, impurity, and grain size on Copper line

resistivity. *Applied Physics Letters* **89**, 113124-113123 (2006).
131. Steinhögl, W., Schindler, G., Steinlesberger, G. & Engelhardt, M. Size-dependent resistivity of metallic wires in the mesoscopic range. *Physical Review B* **66**, 075414 (2002).
132. Steinhögl, W., Schindler, G., Steinlesberger, G., Traving, M. & Engelhardt, M. Comprehensive study of the resistivity of copper wires with lateral dimensions of 100 nm and smaller. *Journal of Applied Physics* **97**, 023706-023707 (2005).
133. McConnell, A.D., Uma, S. & Goodson, K.E. Thermal conductivity of doped polysilicon layers. *Microelectromechanical Systems, Journal of* **10**, 360-369 (2001).
134. Srivastava, G.P. Theory of thermal conduction in nonmetals. *MRS Bulletin*, 445 (2001).
135. Glassbrenner, C.J. & Slack, G.A. Thermal Conductivity of Silicon and Germanium from 3 K to the Melting Point. *Physical Review* **134**, A1058 (1964).
136. Martin, P., Aksamija, Z., Pop, E. & Ravaioli, U. Impact of Phonon-Surface Roughness Scattering on Thermal Conductivity of Thin Si Nanowires. *Physical Review Letters* **102**, 125503-125504 (2009).
137. Hochbaum, A.I. et al. Enhanced thermoelectric performance of rough silicon nanowires. *Nature* **451**, 163-167 (2008).
138. Chen, R. et al. Thermal Conductance of Thin Silicon Nanowires. *Physical Review Letters* **101**, 105501-105504 (2008).
139. Boukai, A.I. et al. Silicon nanowires as efficient thermoelectric materials. *Nature* **451**, 168-171 (2008).
140. Li, B., Wang, L. & Casati, G. Thermal Diode: Rectification of Heat Flux. *Physical Review Letters* **93**, 184301 (2004).
141. Saira, O.-P. et al. Heat Transistor: Demonstration of Gate-Controlled Electronic Refrigeration. *Phys. Rev. Lett.* **99** (2007).
142. Dames, C. Solid-State Thermal Rectification With Existing Bulk Materials. *Journal of Heat Transfer* **131**, 061301-061307 (2009).
143. Jezowski, A. & Rafalowicz, J. Heat flow asymmetry on a junction of quartz with graphite. *Physica Status Solidi (a)* **47**, 229-232 (1978).
144. Marucha, C., Mucha, J. & Rafalowicz, J. Heat flow rectification in inhomogeneous GaAs. *Physica Status Solidi (a)* **31**, 269-273 (1975).
145. Kobayashi, W., Teraoka, Y. & Terasaki, I. An oxide thermal rectifier. *Appl. Phys. Lett.* **95**, 171905 (2009).
146. Yang, N., Zhang, G. & Li, B. Carbon nanocone: A promising thermal rectifier. *Applied Physics Letters* **93**, 243111-243113 (2008).
147. Roberts, N.A. & Walker, D.G. in 11th Intersociety Conference on Thermal and Thermomechanical Phenomena in Electronic Systems (ITHERM) 993-9982008).
148. Chang, C.W., Okawa, D., Majumdar, A. & Zettl, A. Solid-State Thermal Rectifier. *Science* **314**, 1121-1124 (2006).
149. Casati, G. Device physics: The heat is on -- and off. *Nat Nano* **2**, 23-24 (2007).
150. Scheibner, R. et al. Quantum dot as thermal rectifier. *New Journal of Physics*, 083016 (2008).
151. Wang, Z. et al. Ultrafast Flash Thermal Conductance of Molecular Chains. *Science* **317**, 787-790 (2007).
152. Costescu, R.M., Wall, M.A. & Cahill, D.G. Thermal conductance of epitaxial interfaces. *Physical Review B* **67**, 054302 (2003).
153. Lyeo, H.-K. & Cahill, D.G. Thermal conductance of interfaces between highly dissimilar materials. *Physical Review B* **73**, 144301-144306 (2006).
154. Huxtable, S.T. et al. Interfacial heat flow in carbon nanotube suspensions. *Nat Mater* **2**, 731-734 (2003).
155. Panzer, M.A. et al. Thermal Properties of Metal-Coated Vertically Aligned Single-Wall Nanotube Arrays. *Journal of Heat Transfer* **130**, 052401-052409 (2008).
156. Stevens, R.J., Smith, A.N. & Norris, P.M. Measurement of Thermal Boundary Conductance of a Series of Metal-Dielectric Interfaces by the Transient Thermoreflectance Technique. *Journal of Heat Transfer* **127**, 315-322 (2005).
157. Tong, T. & Majumdar, A. Reexamining the 3-omega technique for thin film thermal characterization. *Rev. Sci. Instruments* **77**, 104902 (2006).
158. Yamane, T., Nagai, N., Katayama, S. & Todoki, M. Measurement of thermal conductivity of silicon dioxide thin films using a 3w method. *J. Appl. Phys.* **91**, 9772 (2002).
159. Lee, S.-M. & Cahill, D.G. Heat transport in thin dielectric films. *J. Appl. Phys.* **81**, 2590 (1997).
160. Dames, C. & Chen, G. 1w, 2w, and 3w methods for measurements of thermal properties. *Rev. Sci. Instruments* **76**, 124902 (2005).
161. Kim, E.-K., Kwun, S.-I., Lee, S.-M., Seo, H. & Yoon, J.-G. Thermal boundary resistance at $Ge_2Sb_2Te_5$/ZnS:$SiO_2$ interface. *Appl. Phys. Lett.* **76**, 3864 (2000).
162. Ge, Z., Cahill, D.G. & Braun, P.V. Thermal Conductance of Hydrophilic and Hydrophobic Interfaces. *Physical Review Letters* **96**, 186101-186104 (2006).
163. Gundrum, B.C., Cahill, D.G. & Averback, R.S. Thermal conductance of metal-metal interfaces. *Physical Review B* **72**, 245426 (2005).
164. Swartz, E.T. & Pohl, R.O. Thermal boundary resistance. *Reviews of Modern Physics* **61**, 605 (1989).
165. Reifenberg, J.P. et al. Thermal Boundary Resistance Measurements for Phase-Change Memory Devices. *IEEE Elec. Dev. Lett. (accepted)*, DOI: 10.1109/LED.2009.2035139 (2010).
166. Stoner, R.J. & Maris, H.J. Kapitza conductance and heat flow between solids at temperatures from 50 to 300 K. *Physical Review B* **48**, 16373 (1993).